\documentclass{sig-alternate-05-2015}

\usepackage{epsfig,endnotes}
\usepackage{graphicx,subfigure,array}
\usepackage{times,amsmath,amssymb,verbatim}
\usepackage[usenames,dvipsnames]{color}
\usepackage{cite,footnote,verbatim,threeparttable}

\usepackage[font=small,labelfont=bf,skip=2pt]{caption}
\usepackage{multirow,algorithm,algorithmic}
\usepackage{mdwlist}

\usepackage{hyphenat}
\PassOptionsToPackage{hyphens}{url}
\usepackage[colorlinks,urlcolor=blue,citecolor=Green,breaklinks=true]{hyperref}
\usepackage[hyphenbreaks]{breakurl}
\usepackage{hhline}
\usepackage{booktabs,paralist,dblfloatfix}
\usepackage{footmisc}
\usepackage[table,xcdraw]{xcolor}

\graphicspath{{../Plots/}}

\newcommand{\paragraphb}[1]{\vspace{0.05in}\noindent{\bf\normalsize #1} }

\begin{document}






%

\title{Smartphone Fingerprinting Via Motion Sensors: Analyzing Feasibility at Large-Scale and Studying Real Usage Patterns}
%
%
%
%
%

\numberofauthors{4} 
%
\author{
%
%
\alignauthor Anupam Das\\
       \affaddr{University of Illinois at Urbana-Champaign}\\
       \email{das17@illinois.edu}
\alignauthor  Nikita Borisov\\
       \affaddr{University of Illinois at Urbana-Champaign}\\
       \email{nikita@illinois.edu}
\alignauthor Edward Chou\\
       \affaddr{University of Illinois at Urbana-Champaign}\\
       \email{ejchou2@illinois.edu}
\and  
\alignauthor Muhammad Haris Mughees\\
       \affaddr{Hong Kong University of Science and Technology}\\
       \email{mhmughees@ust.hk}
}

\maketitle

\begin{abstract}
Advertisers are increasingly turning to fingerprinting techniques to track users across the web. As web browsing
activity shifts to mobile platforms, traditional browser fingerprinting techniques become less effective; however, 
device fingerprinting using built-in sensors offers a new avenue for attack. We study the feasibility of using motion sensors to perform device fingerprinting at scale, and explore countermeasures that can be used to protect privacy.

We perform a large-scale user study to demonstrate that motion sensor fingerprinting is effective with 500 users. We also 
develop a model to estimate prediction accuracy for larger user populations; our model provides a conservative estimate of at least 12\% classification accuracy with 100\,000 users. We then investigate the use of motion sensors on the web and find, distressingly, that many sites send motion sensor data to servers for storage and analysis, paving the way to potential fingerprinting. Finally, we consider the problem of developing fingerprinting countermeasures; we evaluate a previously proposed obfuscation technique and a newly developed quantization technique via a user study. We find that both techniques are able to drastically reduce fingerprinting accuracy without significantly impacting the utility of the sensors in web applications.
\end{abstract}

%
%

%
%

%
%




\section{Introduction}

We are in the middle of a war over user privacy on the web. After the failure of the ``Do Not Track'' proposal, users have increasingly started using tools such as ad- and tracker-blocking extensions, as well as private browsing modes, to protect their privacy. In turn, advertisers have started using browser fingerprinting~\cite{Eckersley:2010,canvas-fingerprint,Acar:2014} to track users across the web without the use of cookies. As the battleground shifts to mobile platforms, which are quickly becoming the dominant mode for web browsing~\cite{moretime1,moretime2,moretime3,mobiletraffic}, existing fingerprinting techniques become less effective~\cite{Hupperich:2015,Spooren:2015}; at the same time, new threats emerge: mobile browsers give web pages access to internal motion sensors (accelerometers and gyroscopes) and researchers have showed that imperfections in these sensors can be used to fingerprint \emph{devices}~\cite{anupam:2016,Hupperich:2015,accelprint}, boosting the accuracy of a weakened \emph{browser} fingerprint.  

An important question not addressed by prior work is whether such fingerprinting can be effective at scale, as state-of-the-art techniques~\cite{anupam:2016} have only been evaluated on a set of 100 devices. We first perform a larger-scale evaluation of the methods, collecting motion sensor data from a total of 610 devices, and showing that high accuracy classification is still feasible. We then used the data we collected to develop a model to predict classification accuracy for larger data sets, by fitting a parametric distribution to model inter- and intra-cluster distances. We can then use these distributions to predict the accuracy of a $k$-NN classifier, used with state-of-the-art distance metric learning techniques; our evaluation shows that even with 100\,000 devices, 12--16\% accuracy can be achieved, depending on training set size, which suggests that motion sensor fingerprinting can be effective when combined with even a weak browser fingerprint. Note that because $k$-NN underperforms other classifiers, such as bagged trees, our estimate of accuracy is quite conservative.

A second question we wanted to answer was, how are motion sensors used on the web? We analyzed the static and dynamic JavaScripts used by the Alexa top-100K websites~\cite{Alexa} and identified over 1\,000 instances of motion sensor access. After clustering, we were able to identify a number of common uses of motion sensors, including orientation detection and random number generation. More distressingly, we noted that a large fraction of scripts send motion data back to a server, while others use the presence of motion sensors in advertising decisions. Thus, although we have not been able to identify cases of motion sensor fingerprinting in the wild, the infrastructure for collecting and analyzing this data is already there in some cases. 

These results suggest that motion sensor fingerprinting is a realistic privacy threat. We therefore wanted to understand the feasibility of defending against fingerprinting through browser- or OS-based techniques. Although several defenses have been proposed to mitigate motion sensor fingerprinting, they reduce the potential utility of the sensor data by adding noise or other transforms. We wanted to understand how this trade off between privacy and utility plays out for the likely uses of the device motion API. To do this, we implement a game that uses motion sensors for controls---a relatively demanding application. We then carry out a user study to investigate the impact of privacy protections on the game difficulty. We evaluate an obfuscation method proposed by Das et al.~\cite{anupam:2016} and develop a new quantization-based protection method. Encouragingly, we find that neither method creates a statistically significant impact on motion sensor utility, as measured by both subjective and objective measures. This suggests that user privacy may be preserved without sacrificing much utility.

In summary, we make the following contributions:\nolinebreak
\begin{itemize*}
\item We evaluate the state-of-the-art fingerprinting techniques by Das et al.~\cite{anupam:2016} on a large data set of 610 devices. (\S\ref{sec:realphones})
\item We develop a model for predicting how the $k$-NN classifier will perform on larger data sets and use it to obtain a conservative estimate of fingerprinting accuracy for up to 100\,000 devices. (\S\ref{sec:scalability})
\item We perform a measurement study to evaluate how motion sensor information is used by existing websites. We identify several common uses for motion sensor data and find that motion data is frequently sent to servers. (\S\ref{measurement_study})
\item We develop a new fingerprinting countermeasure that uses quantization of data in polar coordinates. (\S\ref{sec:quantization}) 
\item We carry out a user study to evaluate the impact of our countermeasure, as well as the obfuscation technique proposed by Das et al.~\cite{anupam:2016}, on the utility of motion sensors and find that users experience no significant ill effects from the countermeasures. (\S\ref{sec:user_study})
\end{itemize*}


\paragraphb{Roadmap.} The remainder of this paper is organized as follows. We present background information and related work in Section {\ref{related}}. In Section {\ref{feature_algo}}, we briefly describe our data collection and feature extraction process along with the classification algorithms and metrics used in our evaluation. Section {\ref{dist_distribution}}, describes how we extrapolate fingerprinting accuracy at large scale by deriving intra- and inter-class distance distributions. We present our measurement study on how top websites access motion sensors in Section {\ref{measurement_study}}. Section {~\ref{sec:countermeasures}} evaluates the usability impact of fingerprinting countermeasures through a large-scale online user study. 
Finally, we conclude in Section {~\ref{conclusion}}.

\section{Related Work}{\label{related}}
Fingerprinting devices has been an interesting research area for many decades. It all started with a rich body of research that looked at fingerprinting wireless devices by analyzing the spectral characteristics of wireless transmitters~\cite{Riezenman2000,Li:2006,Patwari:2007}. Researchers then moved onto fingerprinting computers by exploiting their clock skew rate~\cite{Moon99}. Later on, as computers got connected to the Internet, researcher were able to exploit such skew rates to distinguish connected devices through TCP and ICMP timestamps~\cite{Kohno:2005}. Installed software has also been used to track devices, as different devices usually have a different software base installed. Researchers have utilized such strategy to uniquely distinguish subtle differences in the firmwares and device drivers~\cite{Franklin:2006}. Moreover, there are open source toolkits like Nmap~\cite{nmap} and Xprobe~\cite{xprobe} that can fingerprint the underlying operating system remotely by analyzing the responses from the TCP/IP stack. The latest trend in fingerprinting devices is through the web browser. We will now describe some of the most recent and well-known results in this field.

\paragraph{Browser Fingerprinting} The primary application of browser fingerprinting is to uniquely track a user across multiple websites for advertisement purpose. Traditionally this has been done through the injection of cookies. However, privacy concerns have pushed browser developers to provide ways to clear cookies, and also provide options to browse in private mode which does not store long-term cookies. This has forced publishers to come up with new ways to uniquely identify and track users. The Panopticlick project was one of the first works that looked into exploiting easily accessible browser properties such as installed fonts and plug-ins to fingerprint browsers~\cite{Eckersley:2010}. In recent years, researchers have come up with a more advanced technique that uses HTML5 canvas elements to fingerprint the fonts and rendering engines used by the browser~\cite{canvas-fingerprint}. Moreover, users can be profiled and tracked by their browsing history~\cite{olejnik:hal-00747841}. Many studies have shown that all of these techniques are actually used in the wild~\cite{Acar:2013,Acar:2014,nikiforakis:2012}. Researchers have also looked at countermeasures that typically disable or limit the ability of a web publisher to probe particular browser characteristics. Privaricator~\cite{privaricator} is one such approach that adds noise to the fingerprint to break linkability across multiple visits. 

With the rapid growth of smart devices, researchers are now focusing on adopting existing fingerprinting techniques in the context of smart devices. Like cookies, app developers have looked at using device IDs such as Unique Device Identifier (UDID) or International Mobile Station Equipment Identity (IMEI) to track users across multiple applications. However, Apple ceased the use of UDID since iOS 6~\cite{udiddead} and for Andriod accessing IMEI requires explicit user permission~\cite{androidIMEI}. Moreover, due to constrained hardware and software environment existing methods often lack in precision for smartphones and recent studies have shown this to be true~\cite{Hupperich:2015,Spooren:2015}. However, this year Laperdrix et al. have shown that it is in fact possible to fingerprint smartphones effectively through \emph{user-agent string} which is becoming richer every day due to the numerous vendors with their different firmware versions~\cite{Laperdrix:2016}. Others have looked at fingerprinting smartphones by exploiting the personal configuration settings which are often accessible to third party apps~\cite{kurtz:fingerprinting}.

\paragraph{Sensor Fingerprinting} Today's smartphones come with a wide range of sensors, all of which provide different useful functionality. However, such sensors can also provide side-channels that can be exploited by an adversary to uniquely fingerprint smartphones. Recent studies have looked at exploiting microphones and speakers to fingerprint smartphones~\cite{Das:2014,Zhou:2014,BojinovMNB14}.  Others have looked at utilizing motion sensors like accelerometer to uniquely distinguish smartphones~\cite{accelprint,BojinovMNB14}. And most recently, Das et al. have shown that they can improve the fingerprinting accuracy by combining gyroscope with inaudible sound~\cite{anupam:2016}. Our approach builds on the work done by Das et al. However, our work provides a real-world perspective on the problem. We not only show that sensor-based fingerprinting works at large scale but also show how websites are accessing the sensor data in the wild. Moreover, we also provide a new countermeasure technique where we quantize sensor data to lower the resolution of the sensor. We also by perform a large scale user study to where users play a online game to show that our countermeasure does not affect the utility of the sensors.

\section{Features and Evaluation Metrics}{\label{feature_algo}}
In this section we briefly describe the data collection and data preprocessing step. We also discuss the classification algorithms and evaluation metrics used in our evaluation.

\subsection{Data Collection}
\begin{figure}[!h]
\centering
\includegraphics[width=0.75\columnwidth]{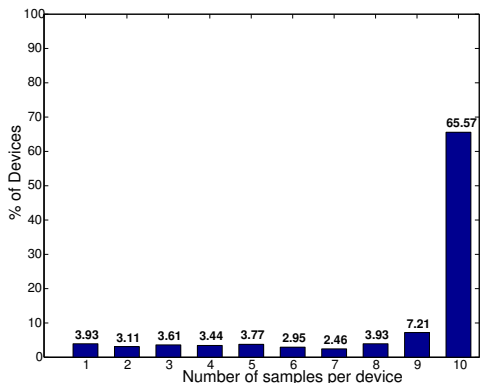}
\caption{Distribution of the number of data samples per smartphone.}
\label{samples_per_device}
\end{figure}

{\label{data_collection}}
To collect sensor data from smartphones we develop a web page\footnote{\url{http://datarepo.cs.illinois.edu/MTurkExp.html}. We obtain IRB approval for collecting sensor data.}. The web page contains a JavaScript to access motion sensors like accelerometer and gyroscope. We create an event listener for device motion in the following manner:\nolinebreak
{\small
\begin{verbatim}
window.addEventListener(`devicemotion',motionHandler)
\end{verbatim}
}
\noindent Once the event listener is registered, the \texttt{motionHandler} function can access accelerometer and gyroscope data in the following 
manner:\nolinebreak
\begin{verbatim}
function motionHandler(event){
  // Access Accelerometer Data
  ax = event.accelerationIncludingGravity.x;
  ay = event.accelerationIncludingGravity.y;
  az = event.accelerationIncludingGravity.z;
  // Access Gyroscope Data
  rR = event.rotationRate;
  if (rR != null){
    gx = rR.alpha;
    gy = rR.beta ;
    gz = rR.gamma;
  }
}
\end{verbatim}
Users are asked to visit our web page while placing their smartphone on a flat surface. Thus, mimicking the scenario where the user has placed his/her smartphone on a desk while browsing a web page. Our web page collects 10 samples consecutively where each sample is 5 seconds worth of sensor data (total participation time is in the range of 1 minute). We found that popular mobile web browsers such as Chrome, Safari and Opera all have a similar sampling rate in the range of 100-120 Hz (Firefox provided a sampling rate close to 180 Hz)\footnote{\url{http://datarepo.cs.illinois.edu/SamplingFreq.html}}. However, the sample rate available at any instance of time depends on multiple factors such as the current battery life and the number of applications running in the background. As a result we received data from participants at various sampling rates ranging from 20 Hz to 120 Hz. Initially, we recruited users through university mass email and social media like Facebook and Twitter. Later on, we recruited participants through Amazon Mechanical Turk~\cite{mturk}. In total, we had a total of 610 participants over a period of three months. We obtained data from 108 different brands (i.e., make and model) of smartphones with different models of iPhone comprising nearly half of the total devices\footnote{We used \url{https://web.wurfl.io/\#learnmore} to obtain the make and model of a smartphone.}. Figure~\ref{samples_per_device} shows the distribution of the different number of samples per device. Since some participation was voluntary for users not using Mechanical Turk, we see that many users provided fewer than 10 samples. 

For the purpose of labeling our data we plant a unique random number inside the cookie. This provides us with ground truth 
data, thus, making it possible to correlate data samples coming from the same physical device.\footnote{%
It is possible that users cleared this cookie, but we do not expect this to happen with enough frequency to significantly affect our data.}

\subsection{Processed Data Streams}{\label{preprocess}}
We process the accelerometer and gyroscope data into four time-series data streams, similar to the way Das et al. do in their 
paper~\cite{anupam:2016}. At any given timestamp, $t$, we have the following two data vectors: 1) acceleration including gravity, 
$\vec{a}(t)=(a_x,a_y,a_z)$ and 2) rotational rate, $\vec{\omega}(t)=(\omega_x,\omega_y,\omega_z)$. The accelerometer value includes gravity, i.e., whenever the device lies stationary flat on top of a surface we get a value of $9.81 ms^{-2}$ along the $z$-axis. To make the fingerprint technique independent of device orientation we take the magnitude of the acceleration vector, $|\vec{a}(t)|=\sqrt{a_x^2+a_y^2+a_z^2}$ as one of our processed data streams. For the gyroscope, since there is no baseline rotational speed (i.e., irrespective of device orientation a stationary device should register $0\,\mathtt{rad}s^{-1}$ rotation rate along all three axes), we consider each axis as a separate source of data stream. Thus, we end up with the following four streams of sensor data: $\{|\vec{a}(t)|,\omega_x(t),\omega_y(t),\omega_z(t)\}$. To obtain frequency domain characteristics we interpolate the non-equally spaced data stream into equally-spaced time-series data by using cubic-spline interpolation.

\subsection{Features}{\label{features}}
Inspired by the most recent work in this field by Das et al.~\cite{anupam:2016}, we extract the same set of 25 features from each data stream. We obtain the feature extraction code base from Das et al~\cite{anupam:2016}. Out of these 25 features, 10 are temporal features and the remaining 15 are spectral features\footnote{A detailed description of each feature is available in the technical report provided by Das et al.~\cite{DasBC15}}. As we have four data streams, we have a total of 100 features to summarize the unique characteristics of the motion sensors.

\subsection{Classification Algorithms and Metrics}{\label{classification-algo}}
\paragraphb{Classification Algorithms:}
Following the approach of Das et al., we use a supervised multi-class classifier. For any supervised algorithm we need to split our data set into training and testing set. The training set (labeled with true device identity) is used to train the classifier while the testing set is used to evaluate how well we can classify unseen data points. In this paper we explore the performance of the following two classifiers: \emph{k}-Nearest Neighbor (\emph{k}-NN) and Random Forest (MATLAB's Treebagger model)~\cite{matlabalgos}; the latter having been found by Das et al.\ to achieve the best classification performance. 

\paragraphb{Evaluation metrics:}
For evaluation metric we use the well-known classification metric \emph{F-score}~\cite{Sokolova2009427}. F-score is the harmonic mean of precision and recall. To compute precision and recall we first compute the true positive ($TP$) rate for each class, i.e., the number of traces that are correctly classified. Similarly, we compute the false positive ($FP$) and false negative ($FN$) as the number of wrongly accepted and wrongly rejected traces, respectively, for each class $i$ ($1\leq i\leq n$). We then compute precision, recall, and F-score for each class using the following 
equations:\nolinebreak
\begin{align}
\mbox{Precision,  } Pr_i &= {TP_i}/(TP_i+FP_i)\\
\mbox{Recall,  } Re_i &= {TP_i}/(TP_i+FN_i)\\
\mbox{F-Score,  } \mathit{F}_i &= ({2\times Pr_i\times Re_i})/(Pr_i+Re_i)
\end{align}
To obtain the overall performance of the system we compute average values across all classes in the following way:\nolinebreak
\begin{align}
\mbox{Avg. Precision,  } \mathit{AvgPr} &= \frac{\sum_{i=1}^{n}Pr_i}{n}\\
\mbox{Avg. Recall,  } \mathit{AvgRe} &= \frac{\sum_{i=1}^{n}Re_i}{n}\\
\mbox{Avg. F-Score,  } \mathit{AvgF} &= \frac{2\times AvgPr\times AvgRe}{AvgPr+AvgRe}
\end{align}
To evaluate our large scale simulation results we use \emph{Accuracy} as our evaluation metric\footnote{\emph{Accuracy} can be thought of as a relaxed version of \emph{F-score}.}. Accuracy is defined as the portion of test traces that are correctly classified.
\begin{align}
\mbox{Accuracy,  } Acc = \frac{\text{\# of samples correctly classified}}{\text{Total test samples}}
\end{align}

\noindent\begin{figure*}[t]
\centering
\begin{tabular}{cccc}
\epsfig{file=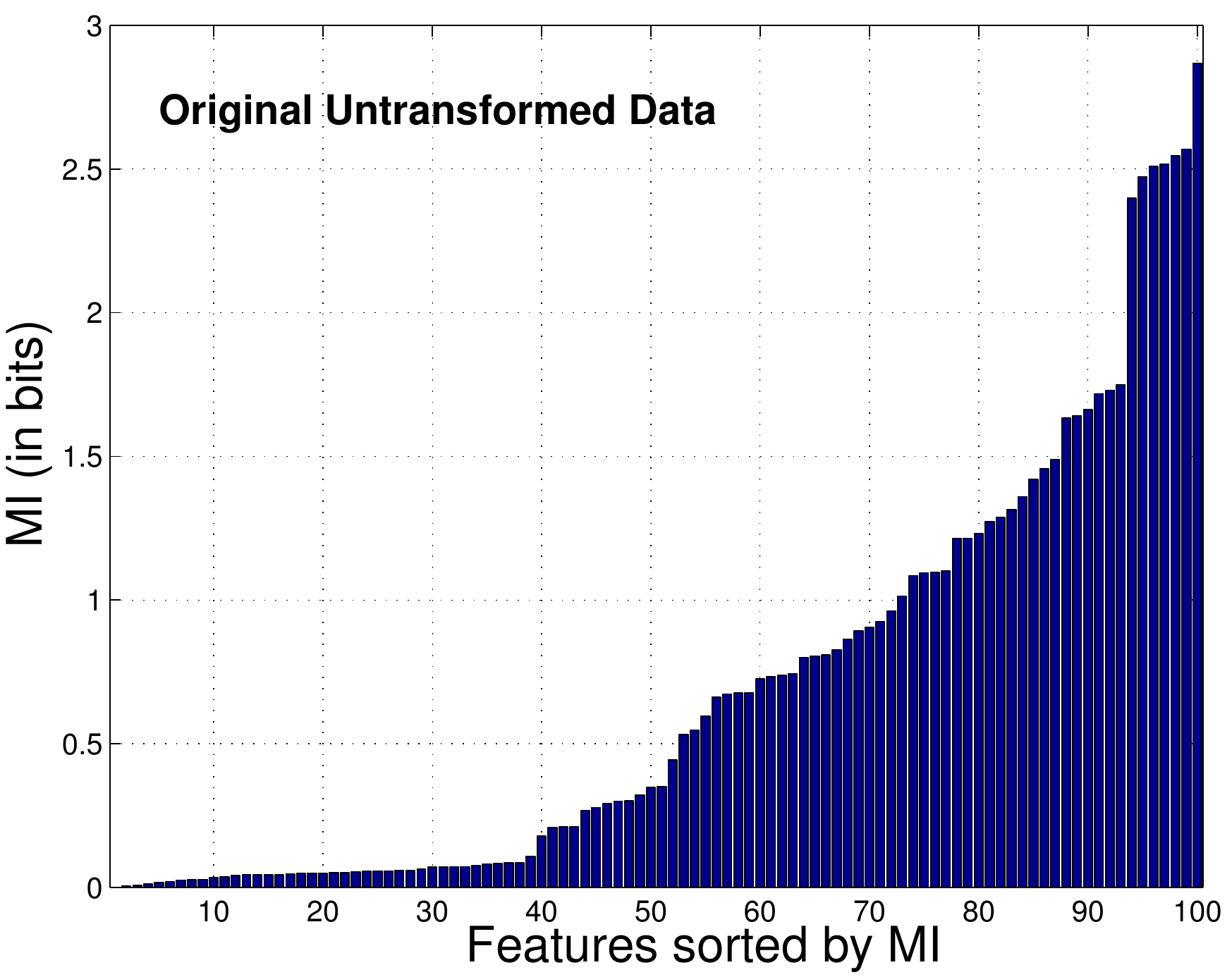,width=0.23\linewidth,clip=}&\epsfig{file=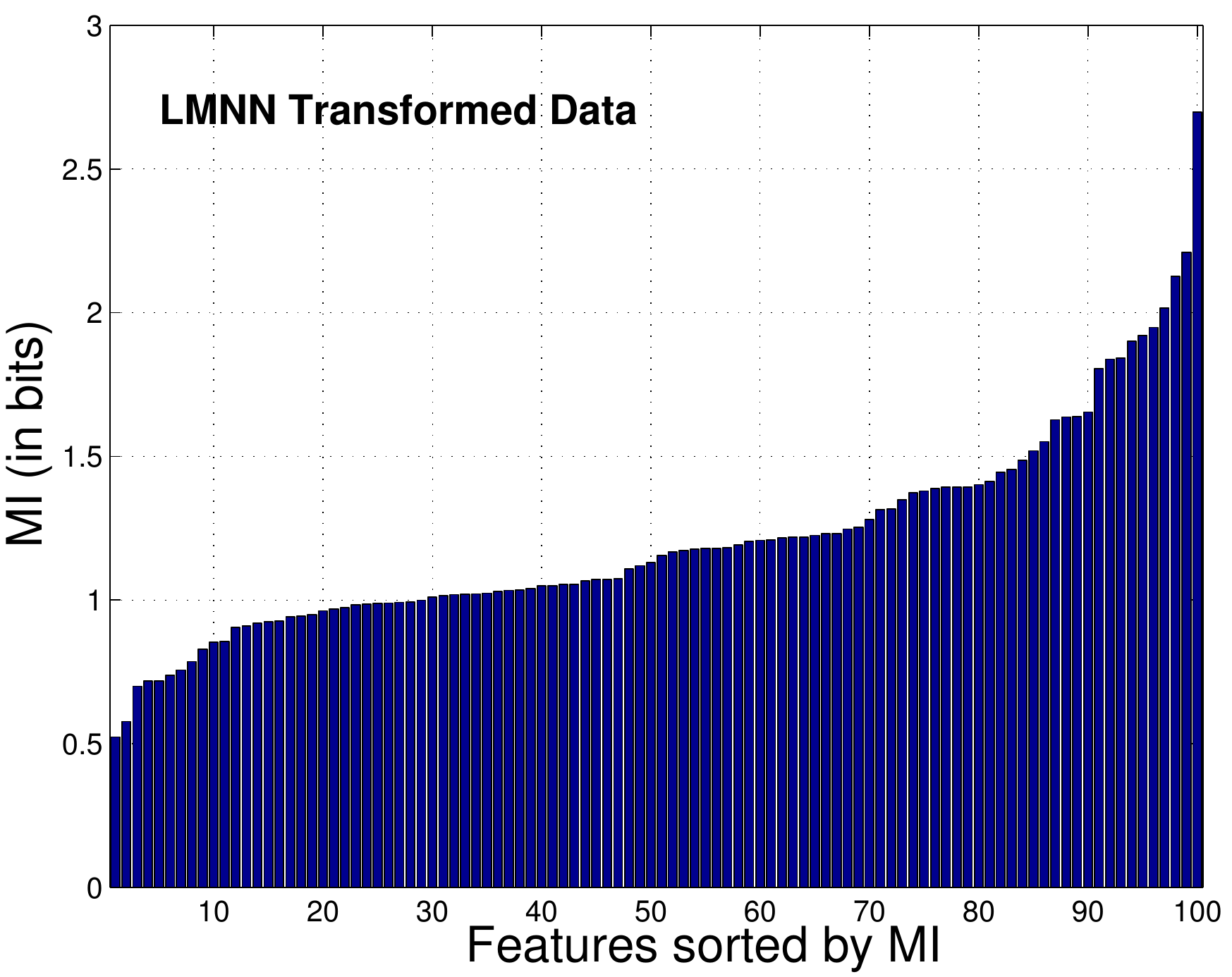,width=0.23\linewidth,clip=}& 
\epsfig{file=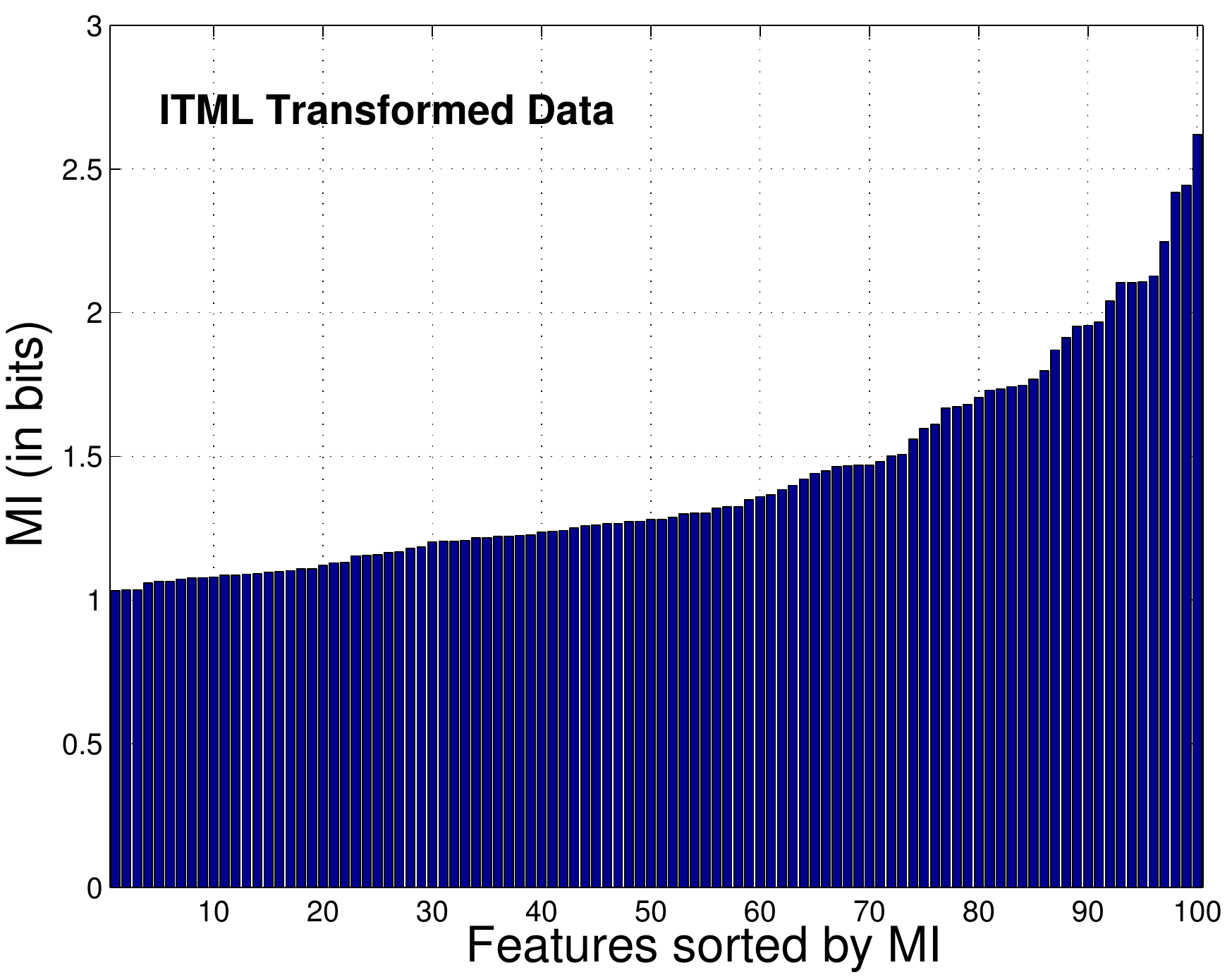,width=0.23\linewidth,clip=}&\epsfig{file=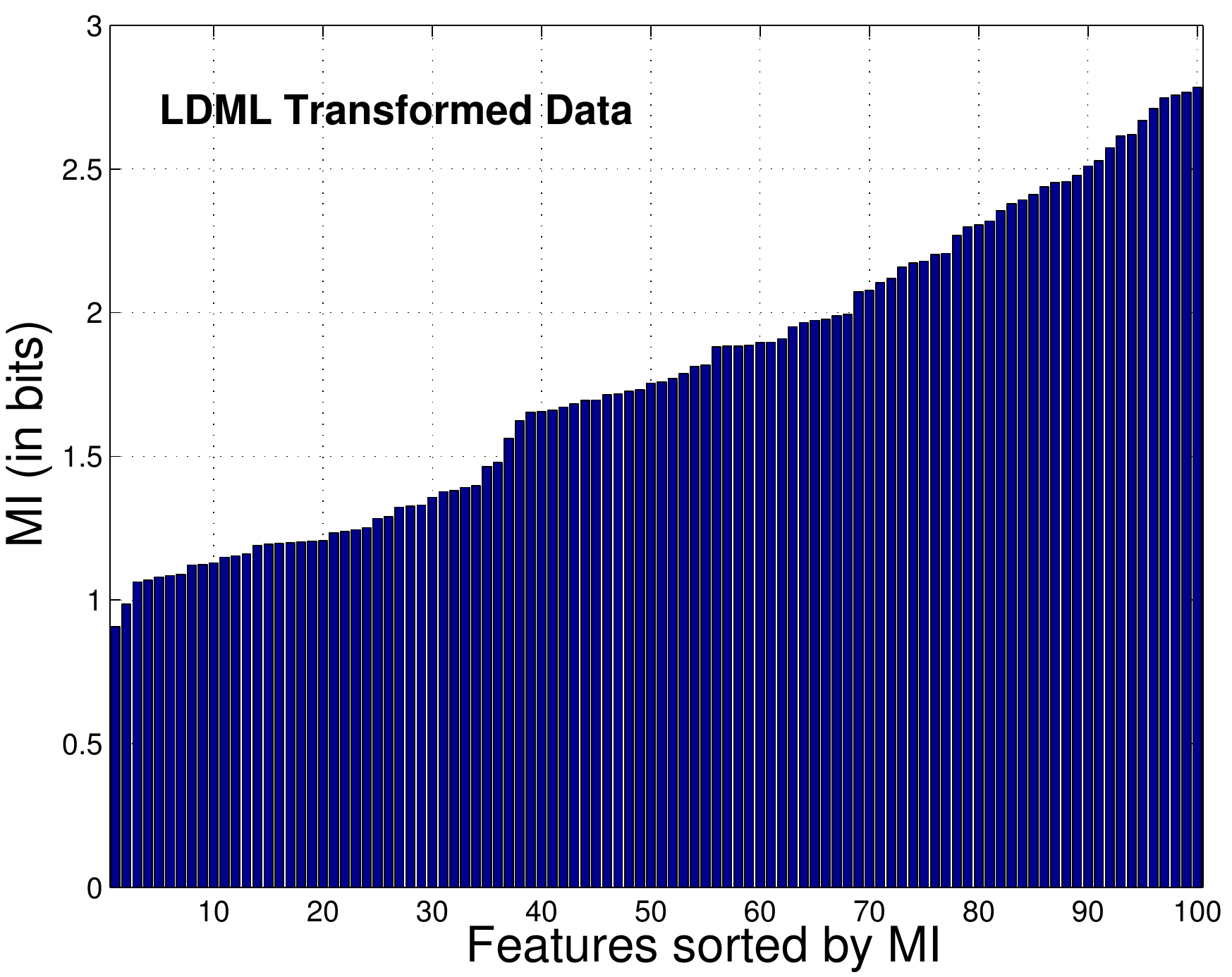,width=0.23\linewidth,clip=}\\
(a)&(b)&(c)&(d)
\end{tabular}
\caption{Comparing mutual information for different metric learning algorithms. Mutual information per feature for (a) untransformed data (b) LMNN transformation (c) ITML transformation, and (d) LDML transformation.} 
\label{Mutual_Info}
\end{figure*}

\section{Fingerprinting Smartphones}{\label{dist_distribution}}
In this section we will first look at how well we can fingerprint our participating smartphones. Next, we will discuss how we can expand our results to simulate experiments with large number of smartphones. Finally, we will provide simulation results for large number of smartphones.

\subsection{Results From Participating Smartphones}
\label{sec:realphones}
We had a total of 610 participants in our data collection study. To evaluate the performance of the classifiers, we first split our data set in training and testing set. As we have devices with different number of data samples (see figure~\ref{samples_per_device}), we evaluate F-score for different size of training set. We \emph{randomly} choose the training and testing samples. To prevent any bias in the selection of the training and testing set we rerun our experiments 10 times and report the average F-score \footnote{We also compute the 95\% confidence interval for F-score, but we found it to be less than 1\% in most cases.}. Table~\ref{realworld_result} summarizes the average F-score for different number of training samples per device. 

\noindent\begin{minipage}{1.0\columnwidth}
\centering 
\captionof{table}{Average F-score for different training set size} 
\resizebox{0.65\columnwidth}{!}{
\begin{tabular}{|c|c|c|}
\hline
{Training}&{Number}&{Avg. F-score (\%)}\\
\cline{3-3}
{samples}&{of}&Random\\
{per device}&{devices}&Forest\footnote{\scriptsize{100 bagged decision trees}}\\
\hline
1&586&33\\
\hline
2&567&65\\
\hline
3&545&78\\
\hline
4&524&83\\
\hline
5&501&86\\
\hline
6&483&88\\
\hline
7&468&89\\
\hline
8&444&89\\
\hline
9&400&90\\
\hline
\end{tabular}}
\label{realworld_result}
\end{minipage}

From Table~\ref{realworld_result} we see that we can achieve high classification accuracy even for this larger data set. 
With five training samples, which correspond to about 25\,s of data, accuracy is 86\%, increasing to 90\% with 9 training samples. Even with a single 5\,s sample, we obtain 33\% accuracy, which may be sufficient if a small amount of extra information can be obtained through other browser fingerprinting techniques, however weak.

In terms of performance, we found that on average it takes around 100--200 ms to match a new fingerprint to an existing fingerprint. For our experiments we use a desktop machine with an Intel i7-2600 3.4GHz processor with 12GiB RAM.

\subsection{Analyzing Scalability}
\label{sec:scalability}
Although we have shown that we can reliably fingerprinting a few hundred devices, in real-world scenarios the fingerprinted population will be much larger. It is not feasible for us to collect data on much larger data sets; instead, we develop a model to predict how well a classifier will perform as the number of devices grows. However, although random forest provides the best classification performance, on our data set, its operation is hard to model, as different trees use a different random sample of features. We therefore base our analysis on nearest-neighbor classifier ($k$-NN), which uses a distance metric that we can model parametrically. Note that $k$-NN does not perform as well as random forest; as a result, our estimates are a \emph{conservative} measure of the actual attainable classification accuracy.

\noindent\begin{figure*}[b]
\centering
\begin{tabular}{cccc}
\epsfig{file=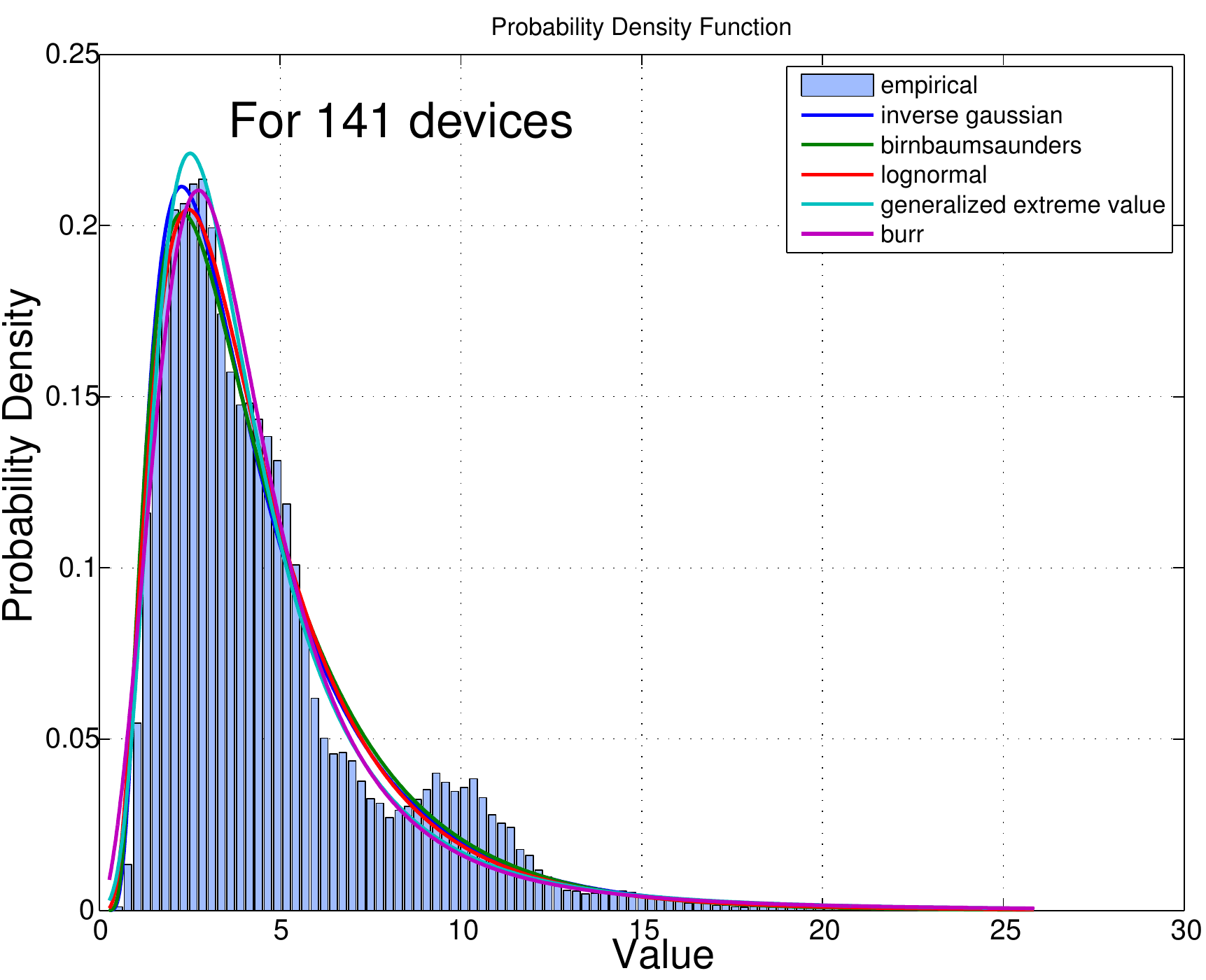,width=0.23\linewidth,clip=}&\epsfig{file=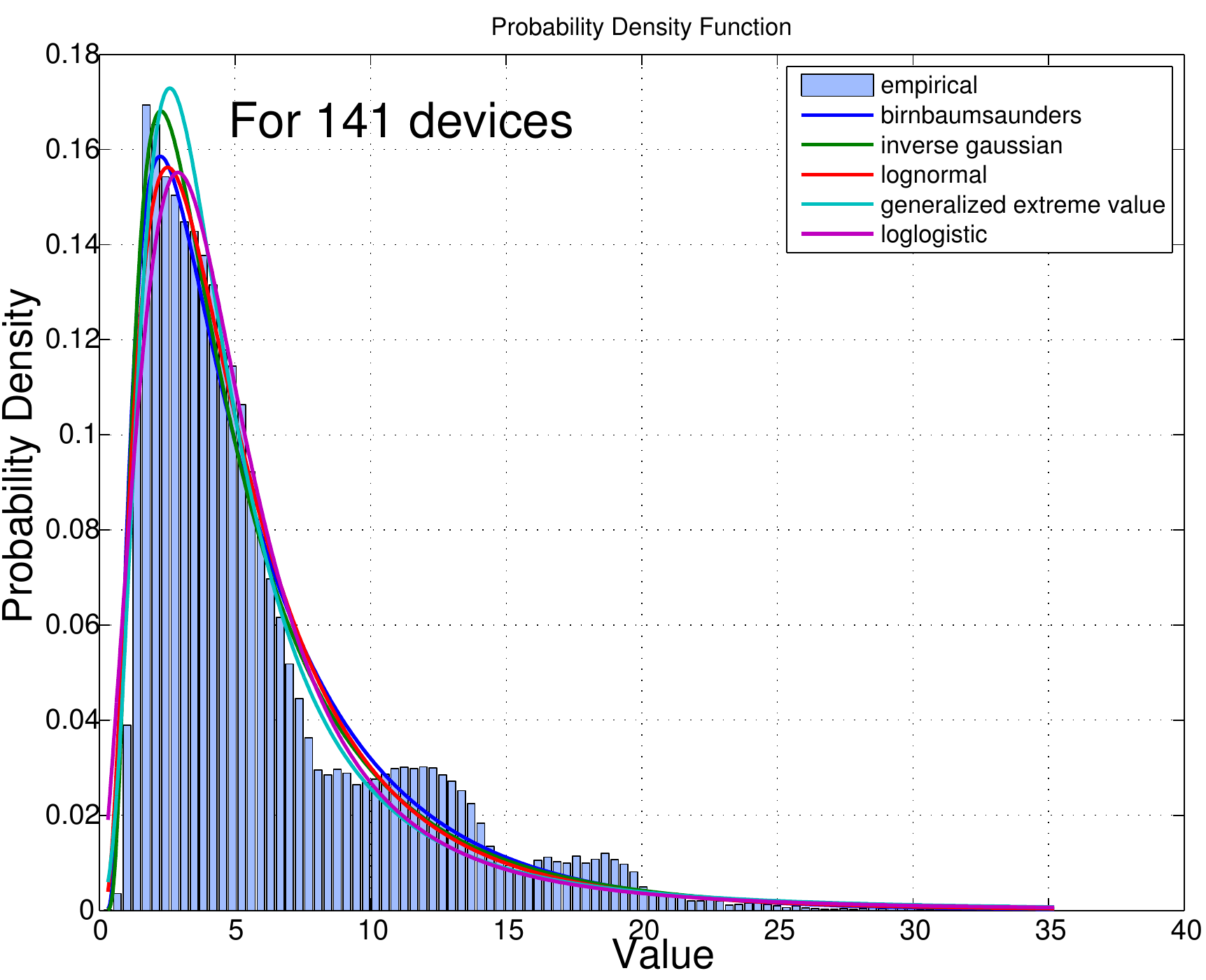,width=0.23\linewidth,clip=}& 
\epsfig{file=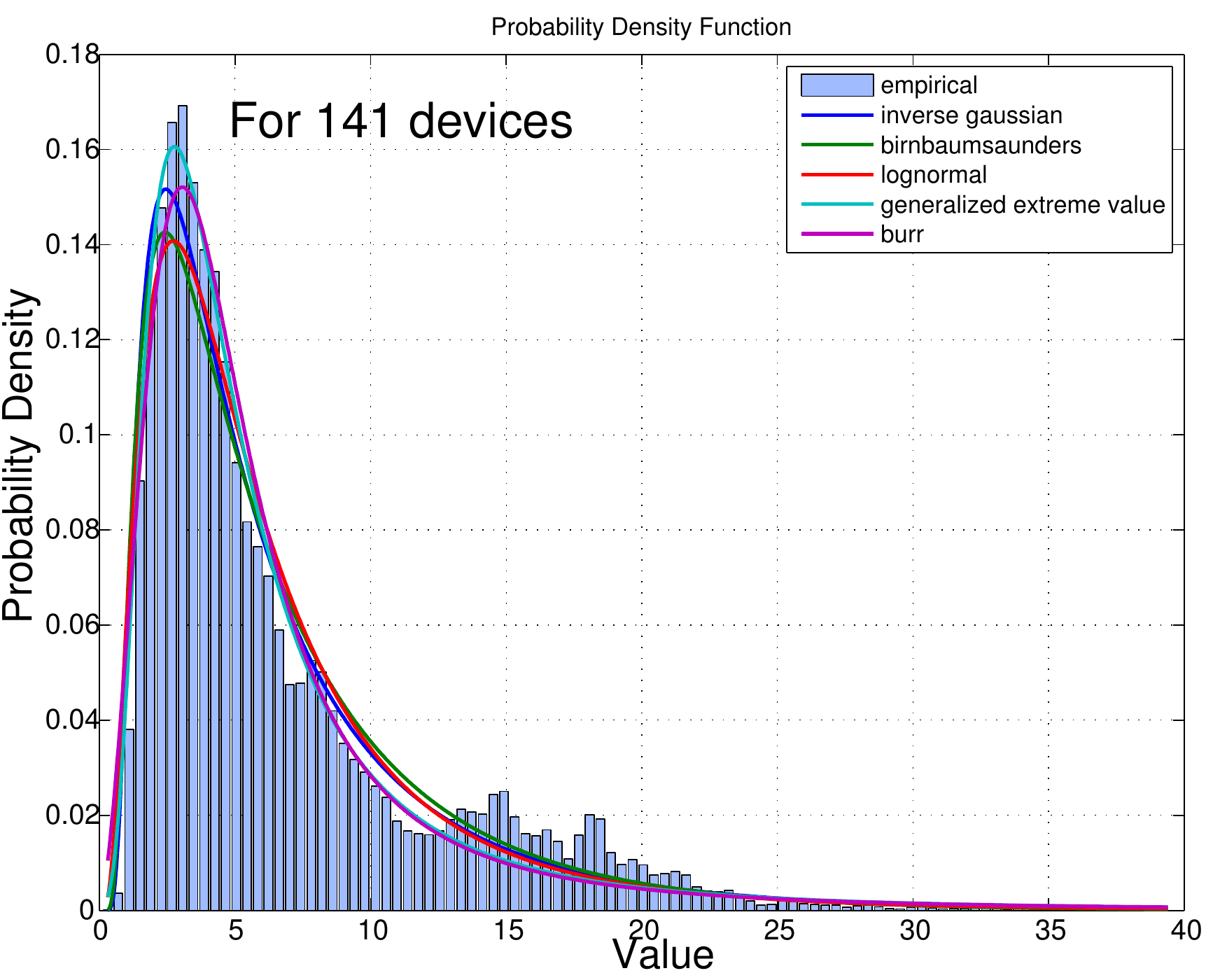,width=0.23\linewidth,clip=}&\epsfig{file=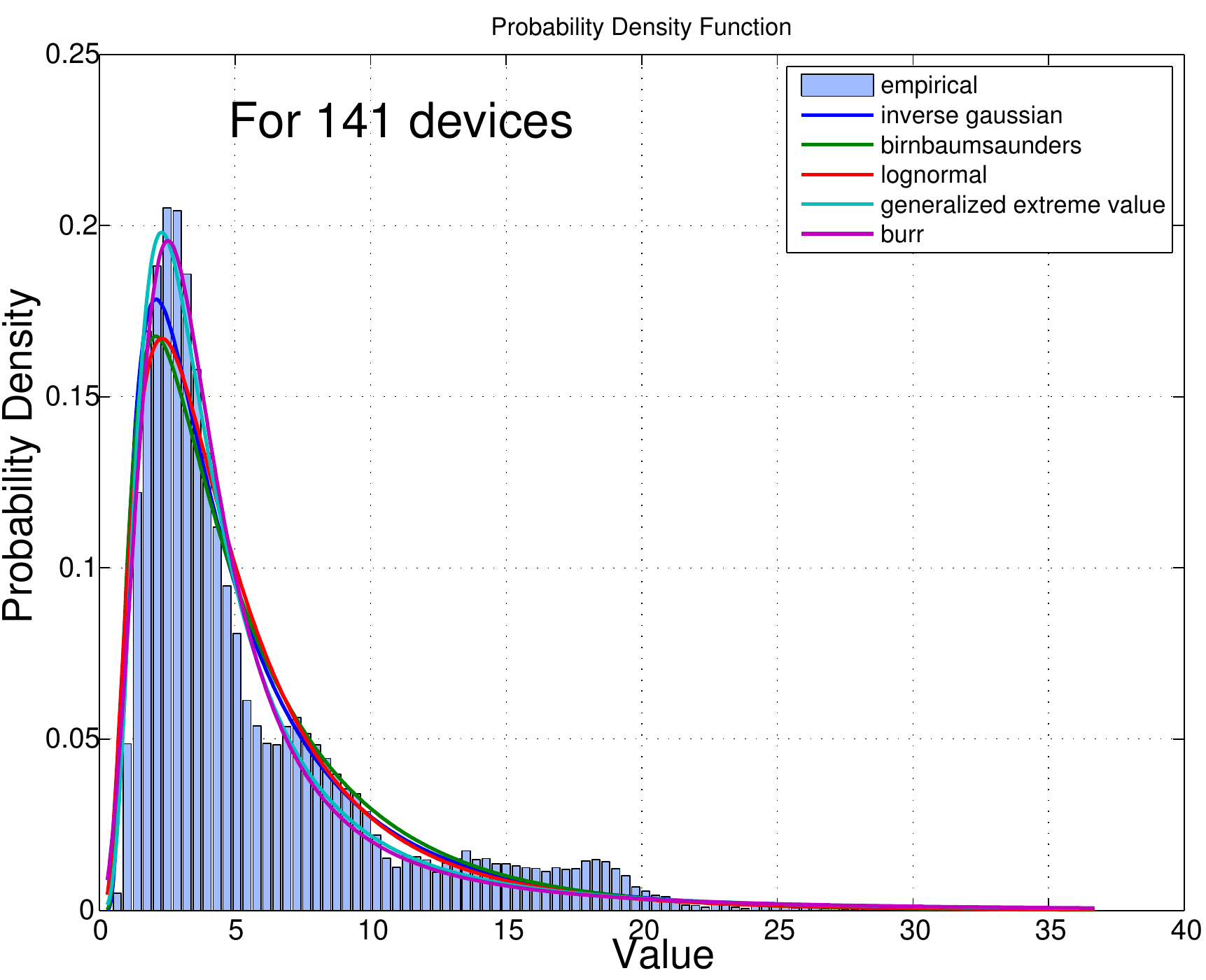,width=0.23\linewidth,clip=}
\end{tabular}
\caption{Estimated inter-device distance distributions for 4 subsets of devices where each subset contains 141 devices.} 
\label{diff_device_distr}
\end{figure*}

\subsubsection{Distance Metric Learning}
The $k$-NN algorithm relies on a distance metric to identify neighboring points. It is possible to compute simple Euclidean distance between feature vectors; however, this is unlikely to yield optimal results as some features will tend to dominate. 
Learning a better distance (or similarity) metric between data points has received much attention in the field of machine learning, pattern recognition and data mining for the past decade~\cite{BelletHS13}. Handcrafting a good distance metric for a specific problem is generally difficult and this has led to the emergence of metric learning. The goal of a distance metric learning algorithm is to take advantage of prior information, in form of labels, to automatically learn a transformation for the input feature space. A particular class of distance function that exhibits good generalization performance for distance-based classifiers such as \emph{k}-NN, is Mahalanobis metric learning~\cite{Kostinger:2012}. The aim is to find a global, linear transformation of the feature space such that relevant dimensions are emphasized while irrelevant ones are discarded. The linear transformation performs arbitrary rotations and scalings to conform to the desired geometry. After projection, Euclidean distance between data points is measured.

State-of-the-art Mahalanobis metric learning algorithms include  \emph{Large Margin Nearest Neighbor} (LMNN)~\cite{weinberger2009distance}, \emph{Information Theoretic Metric Learning} (ITML)~\cite{Davis:2007} and \emph{Logistic Discriminant Metric Learning} (LDML)~\cite{Guillaumin:2009}. A brief description of these metric learning algorithms is provided by K\"{o}stinger et al.~\cite{Kostinger:2012}. To understand how these metric learning algorithms improve the performance \emph{k}-NN classifier, we first plot the \emph{mutual information} (MI) of each feature before and after each transformation. Figure~\ref{Mutual_Info} shows the amount of mutual information per feature under both untransformed and transformed settings.

\noindent\begin{minipage}{1.0\columnwidth}
\centering 
\captionof{table}{Performance of different metric learning algorithms} 
\resizebox{0.65\columnwidth}{!}{
\begin{tabular}{|c|c|c|c|}
\hline
\multicolumn{4}{|c|}{Avg. F-score for \emph{k}-NN\footnote{\scriptsize{$k=1$, 3 training samples per device}}}\\
\hline
{Untransformed}&{LMNN}&{ITML}&{LDML}\\
\hline
35&41&46&50\\
\hline
\end{tabular}}
\label{metric_performance}
\end{minipage}

Figure~\ref{Mutual_Info} shows a clear benefit of the distance metric learning algorithms. All of the transformations provide higher degree of mutual information compared to the original untransformed data. Among the three transformations we see that LDML on average provides slightly higher amount mutual information per feature. This is confirmed when we rerun the \emph{k}-NN classifier on the transformed feature space. Table~\ref{metric_performance} highlights the average F-score for different metric learning algorithms. We see that for our data set, LDML seems to be the best choice. We, therefore, use LDML algorithm to transform our feature space before applying \emph{k}-NN for the rest of the paper. However, even with LDML, $k$-NN underperforms random forest, as seen in Table~\ref{realworld_KNN_result}: our $F$-score drops from 78\% to 54\% with 3 samples and from 86\% to 64\% with 5 samples.

\noindent\begin{minipage}{1.0\columnwidth}
\centering 
\captionof{table}{Average F-score of $k$-NN after LDML} 
\resizebox{0.9\columnwidth}{!}{
\begin{tabular}{|c|c|c|c|c|}
\hline
{Training}&{Number}&\multicolumn{3}{c|}{Avg. F-score (\%)}\\
\cline{3-5}
{samples}&{of}&\multirow{2}{*}{\emph{k}-NN}\footnote{\scriptsize{$k=1$}}&\multirow{2}{*}{\emph{k}-NN+LDML}\footnote{\scriptsize{$k=1$}}&Random\\
{per device}&{devices}&&&Forest\footnote{\scriptsize{100 bagged decision trees}}\\
\hline
1&586&24&38&33\\
\hline
2&567&31&43&65\\
\hline
3&545&35&50&78\\
\hline
4&524&36&52&83\\
\hline
5&501&38&54&86\\
\hline
6&483&38&54&88\\
\hline
7&468&38&53&89\\
\hline
8&444&37&52&89\\
\hline
9&400&35&50&90\\
\hline
\end{tabular}}
\label{realworld_KNN_result}
\end{minipage}

\noindent\begin{figure*}[t]
\centering
\begin{tabular}{ccc}
\epsfig{file=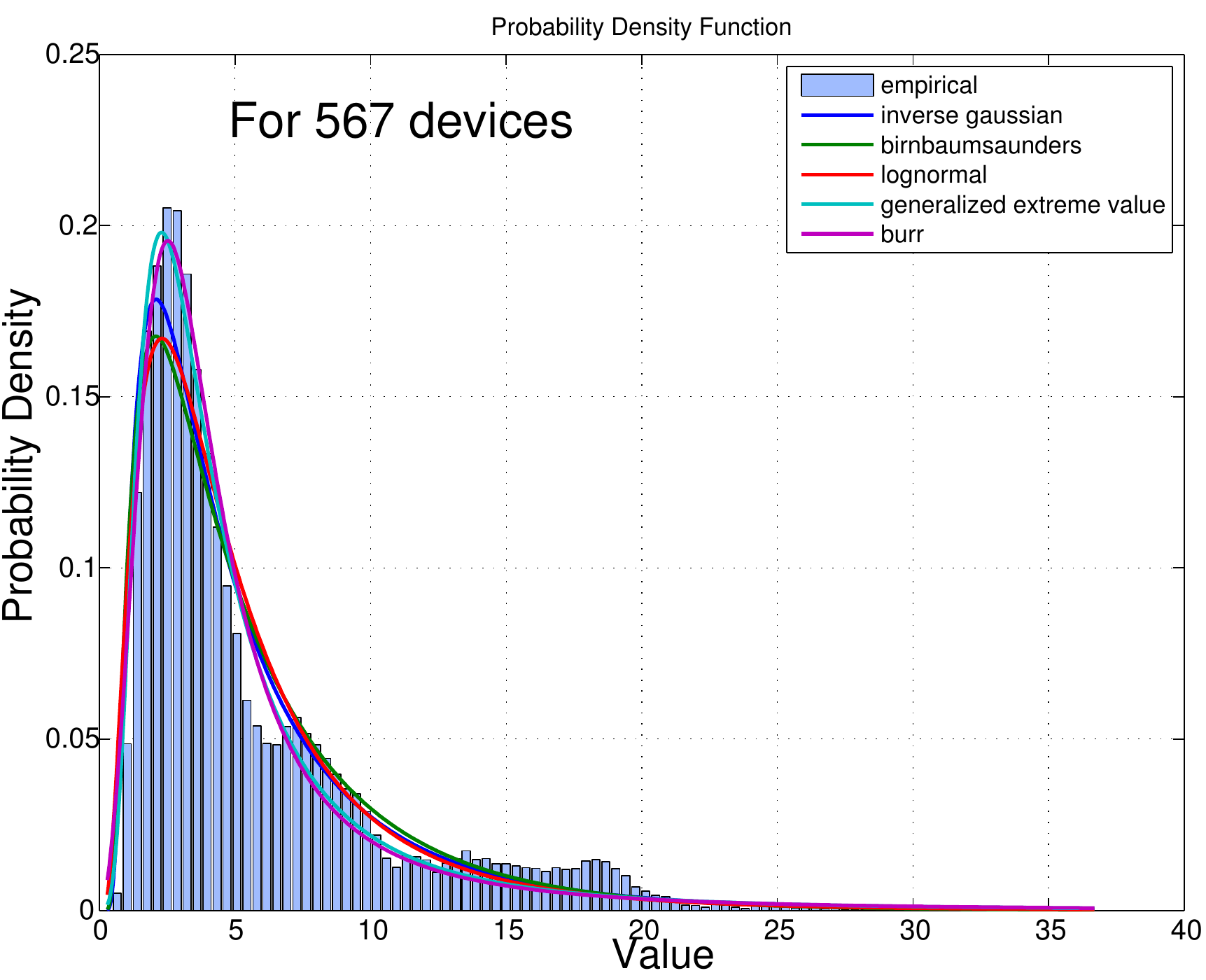,width=0.32\linewidth,clip=}&\epsfig{file=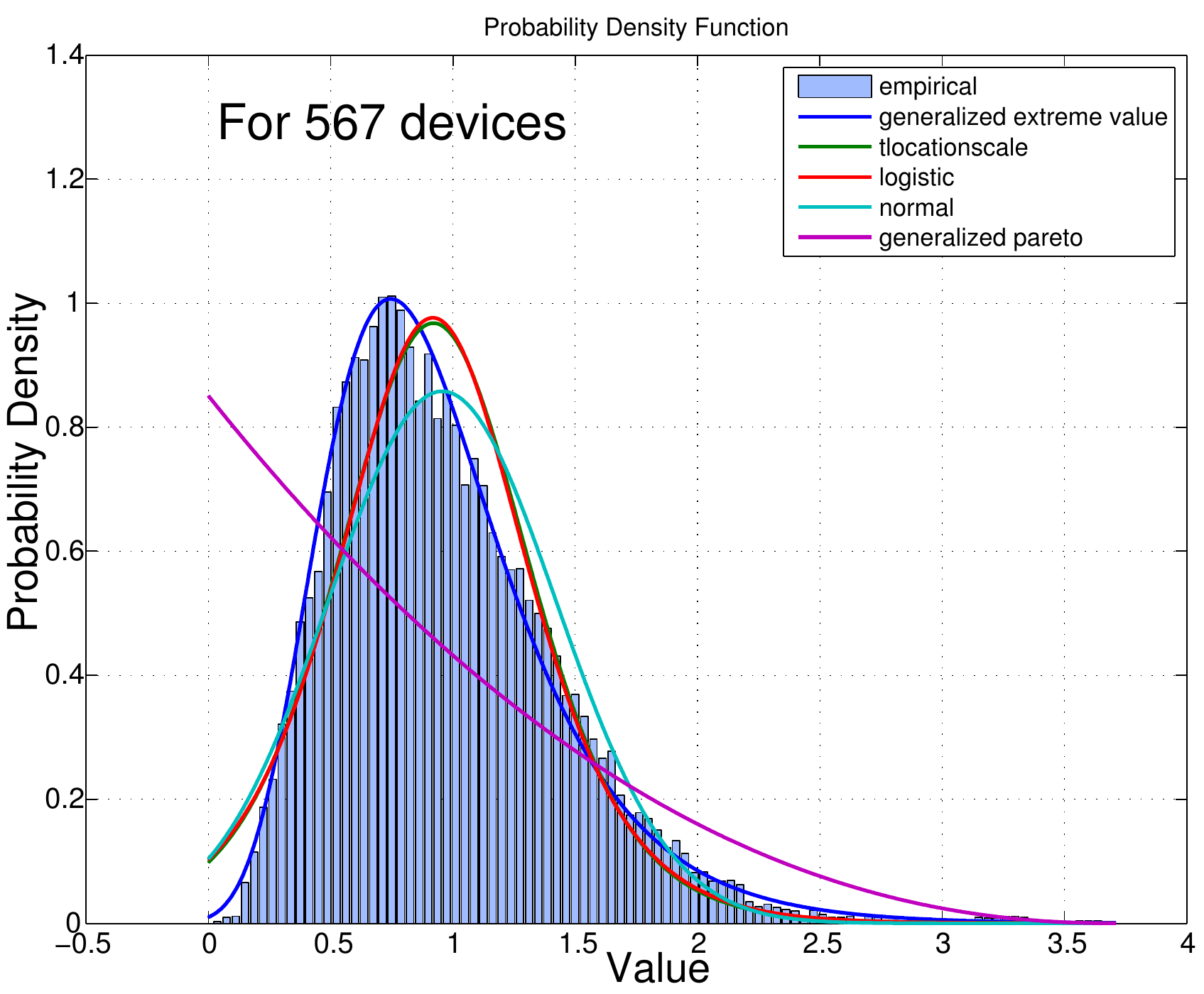,width=0.32\linewidth,clip=}& 
\epsfig{file=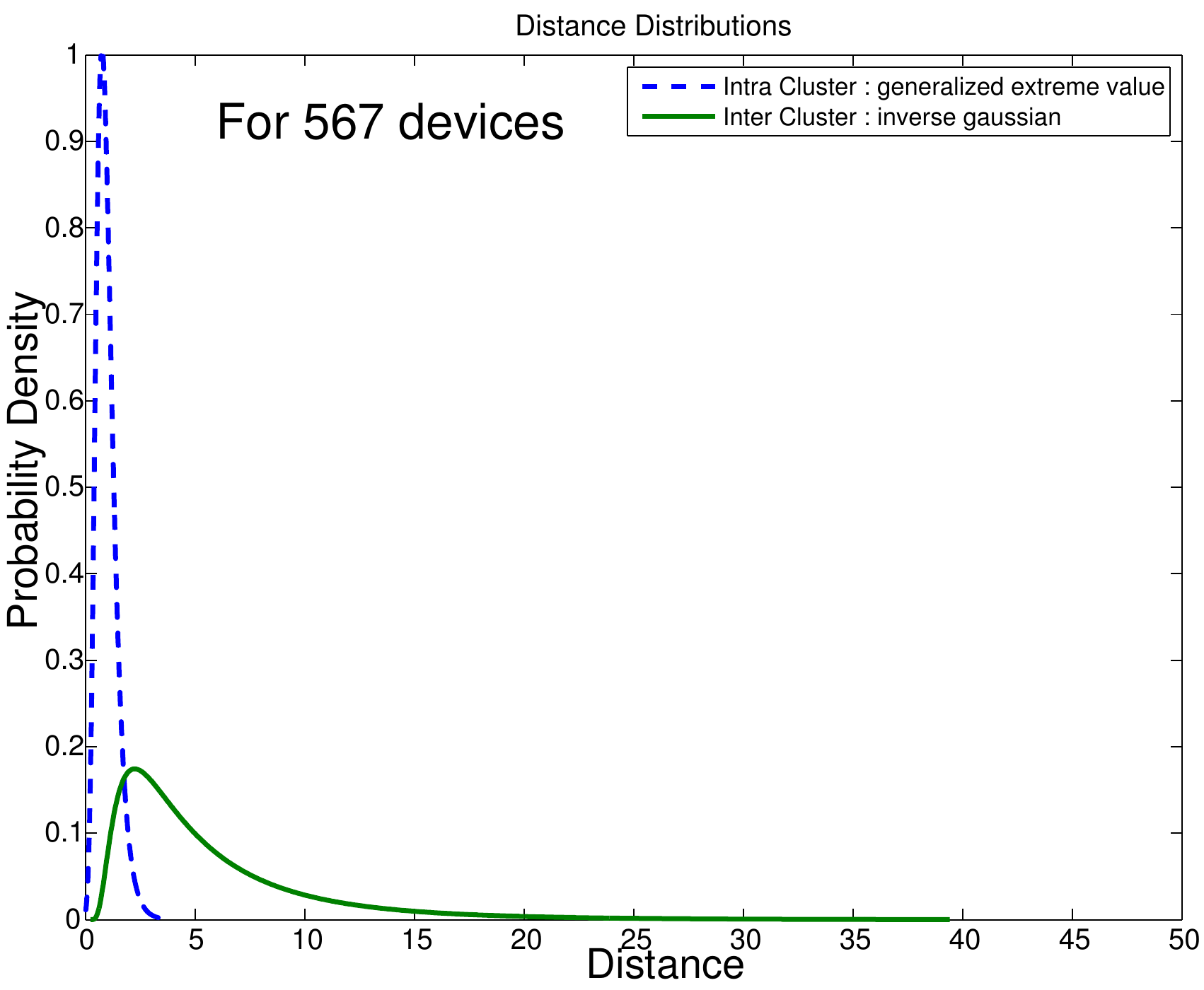,width=0.32\linewidth,clip=}\\
(a)&(b)&(c)
\end{tabular}
\caption{Estimated distributions for (a) inter-device distance ($C_{inter}$) (b) intra-device distance ($C_{intra}$). (c) Difference between intra and inter-device distance distribution.} 
\label{dist_distr}
\end{figure*}

\subsubsection{Intra and Inter-Device Distance}
To predict how $k$-NN will operate on larger data sets, we proceed to derive a distribution for distances between samples from different devices (inter-device), and a second distribution for distances between different samples from the same device (intra-device), after first applying the LDML transformation to the feature space. Since each data sample is a point in a $n$-dimensional feature space, we compute the Euclidean distance between any two data samples using the the following equation:\nolinebreak
\begin{equation}
d(p,q) = \sqrt{\sum_{i=1}^{n}(p_i-q_i)^2}
\end{equation}
where $p$ and $q$ represent two feature vectors defined as follows, $p=(p_1,p_2,...,p_n)$, $q=(q_1,q_2,...,q_n)$. We then group distances between feature vectors from the same device into one class $C_{intra}$ and distances between feature vectors from different devices into another class $C_{inter}$. Class $C_{intra}$ and $C_{inter}$ can be defined as follows:
\nolinebreak
\begin{align}
C_{intra}&=\{ x : x=d(p,q), p\in{D_i}, q\in{D_i}, \forall{i}\in{D}\}\nonumber\\
C_{inter}&=\{ x : x=d(p,q), p\in{D_i}, q\in{D_j}, i\neq{j}, \forall{i,j}\in{D}\}\nonumber
\end{align}
where $D$ refers to the set of all devices (we consider only devices with at least 2 traing samples, we have 567 such devices). We can now fit an individual distribution for each class. To do this we utilize MATLAB's \emph{fitdist} function~\cite{fitdist}. To avoid overfitting, we distribute our devices into four equal subsets. We then fit and compare distributions from each subset. Figure~\ref{diff_device_distr} shows the top five estimated inter-device distance ($C_{inter}$) distributions for each subset of devices. Here, the distributions are ranked based on \emph{Akaike Information Criterion} (AIC)~\cite{Akaike1998}. From figure~\ref{diff_device_distr} we can see that the top five distributions are more or less consistent across all four subsets. 

Next, we plot the same inter-device distance distribution but this time we consider data from all 567 devices. Figure~\ref{dist_distr}(a) highlights the top five distributions. Comparing figure~\ref{diff_device_distr} and figure~\ref{dist_distr}(a), we see that the most representative inter-device distance distribution is an \emph{Inverse Gaussian} distribution. Similarly, we find that the most likely intra-device distance distribution ($C_{intra}$) is a \emph{Generalized extreme value} distribution as shown in figure~\ref{dist_distr}(b). Figure~\ref{dist_distr}(c) shows the difference between intra and inter-device distance distribution.

\begin{algorithm}[!h]
\caption{Simulating \emph{k}-NN classifier}
\label{simulate_knn}
\begin{algorithmic}
\STATE  {\bf Input:} {$k$, $N$, $D$, $Distr_{intra}$, $Distr_{inter}$, $Runs$}
\STATE{$\hspace{24pt}$ $k$ -- number of nearest neighbors (odd integer)}
\STATE{$\hspace{24pt}$ $N$ -- number of training samples per device}
\STATE{$\hspace{24pt}$ $D$ -- number of devices}
\STATE{$\hspace{24pt}$ $Distr_{intra}$ -- intra-device distance distribution}
\STATE{$\hspace{24pt}$ $Distr_{inter}$ -- inter-device distance distribution}
\STATE{$\hspace{24pt}$ $Runs$ -- number of runs}
\STATE {\bf Output:} {$Acc$}
\STATE{$\hspace{24pt}$ $Acc$ -- Average classification accuracy}
\STATE{$d \leftarrow \{\}$ \#list of (distance,label) tuple}
\STATE{$Acc\leftarrow 0$}
\FOR{$i:=1$ to $Runs$}
	\STATE{\#add $N$ intra-distances and label each with 0}
	\FOR{$j:=1$ to $N$}
    	\STATE{$d \leftarrow d + \{(random(Distr_{intra}),0)\}$}
	\ENDFOR
	\STATE{\#add $N{\times}(D-1)$ inter-distances and label each with 1}
	\FOR{$j:=1$ to $N{\times}(D-1)$}
    	\STATE{$d \leftarrow d + \{(random(Distr_{inter}),1)\}$} 
	\ENDFOR
	\STATE{$d \leftarrow sort(d)$ \#in ascending order of distance}
	\STATE{$l \leftarrow label(d,k)$ \#return label for top $k$ elements}
	\STATE{$imposters \leftarrow sum(l)$ \#sum top $k$ labels}
	\IF{$imposters < k/2$}
		\STATE{$Acc\leftarrow Acc+1$ \#correct decision}
	\ENDIF
\ENDFOR
\STATE { $Acc\leftarrow Acc/Runs$}
\STATE {\textbf{return} $Acc$}
\end{algorithmic}
\end{algorithm}

\subsubsection{Simulating A Large Number Of Smartphones}
Now that we have representative distributions for intra and inter-device distances, we can simulate a \emph{k}-NN classifier. The pseudo code for simulating \emph{k}-NN classifier is provided in Algorithm~\ref{simulate_knn}. The algorithm works as follows. Let us assume that there are $D$ devices and for each device we have $N$ training samples. Now, for any given test sample, a \emph{k}-NN classifier, first computes $N{\times}D$ distances of which $N$ distances are with samples from the same device and $N{\times}(D-1)$ distances are with all samples belonging to $(D-1)$ other devices. We emulate these distances by drawing $N$ and $N{\times}(D-1)$ distances from our representative intra and inter-device distance distributions, respectively. \emph{k}-NN classifier then inspects the class label for the $k$ nearest neighbors. We can emulate this step by sorting the distances and picking the $k$ lowest distances. Lastly, \emph{k}-NN classifier outputs the class label with the majority vote. To emulate this step we assign each distance a label of either $0$ (meaning distance from same device) or $1$ (meaning distance from different device). We then check if label-$0$ dominates over label-$1$, if so we count that as a successful classification. This whole process repeats multiple times to provide us with an average classification accuracy.

\noindent\begin{figure*}[t]
\centering
\begin{tabular}{ccc}
\epsfig{file=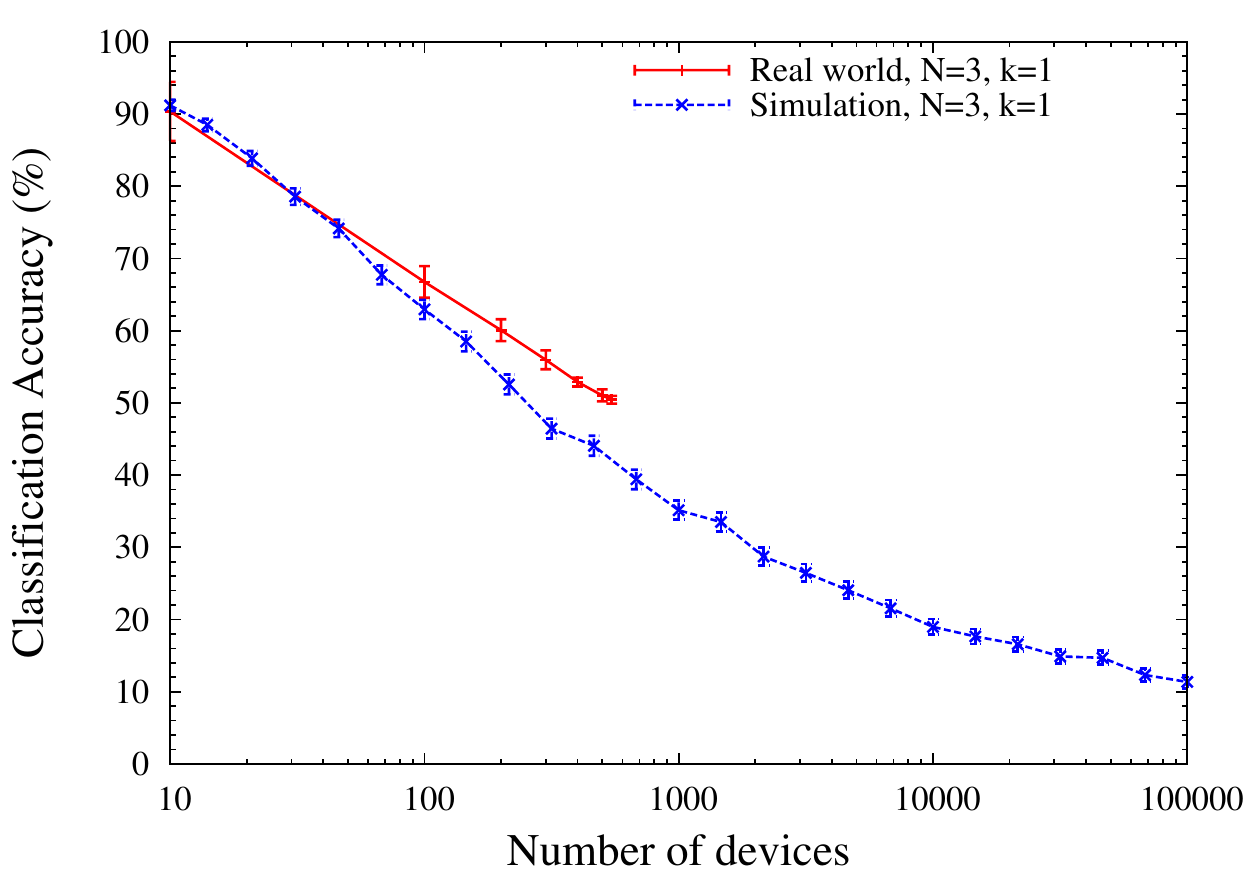,width=0.32\linewidth,clip=}&\epsfig{file=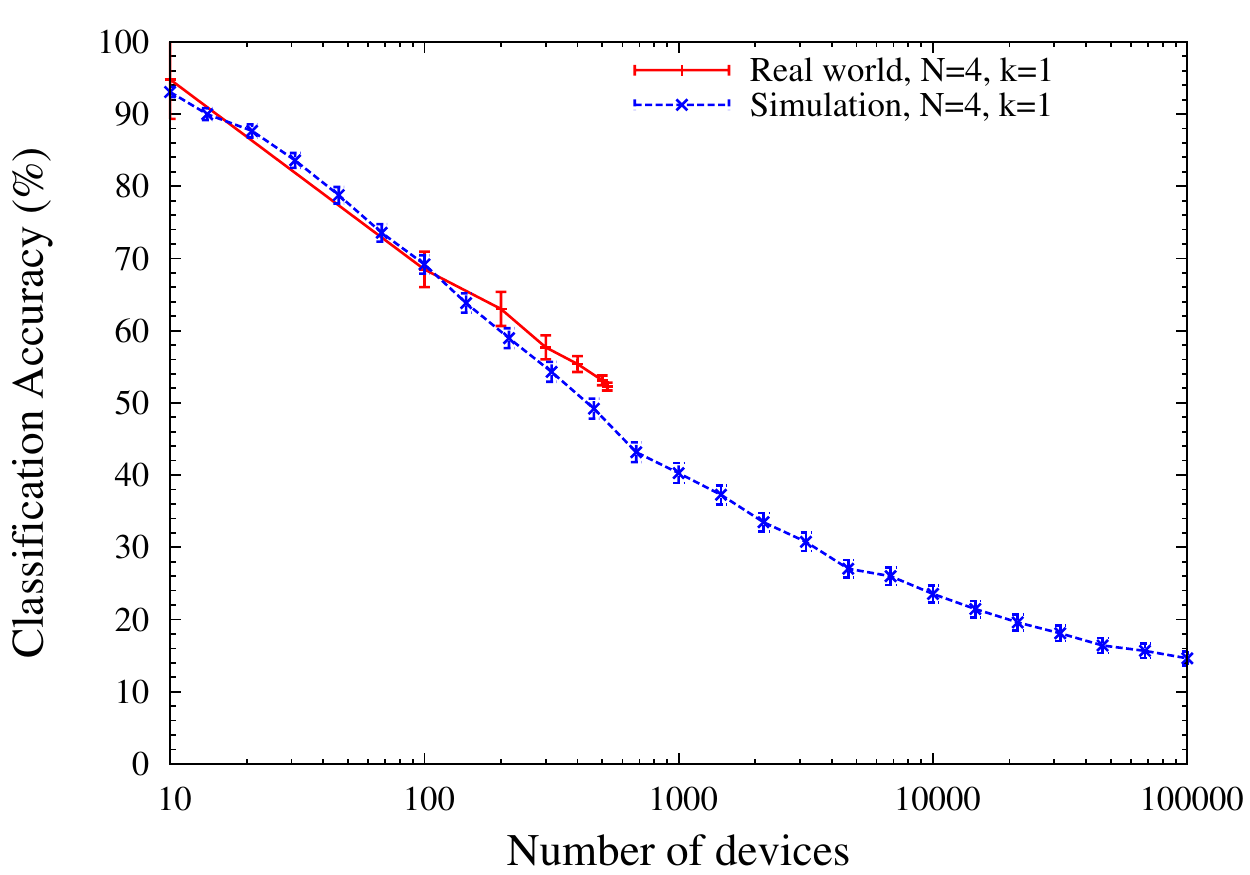,width=0.32\linewidth,clip=}&\epsfig{file=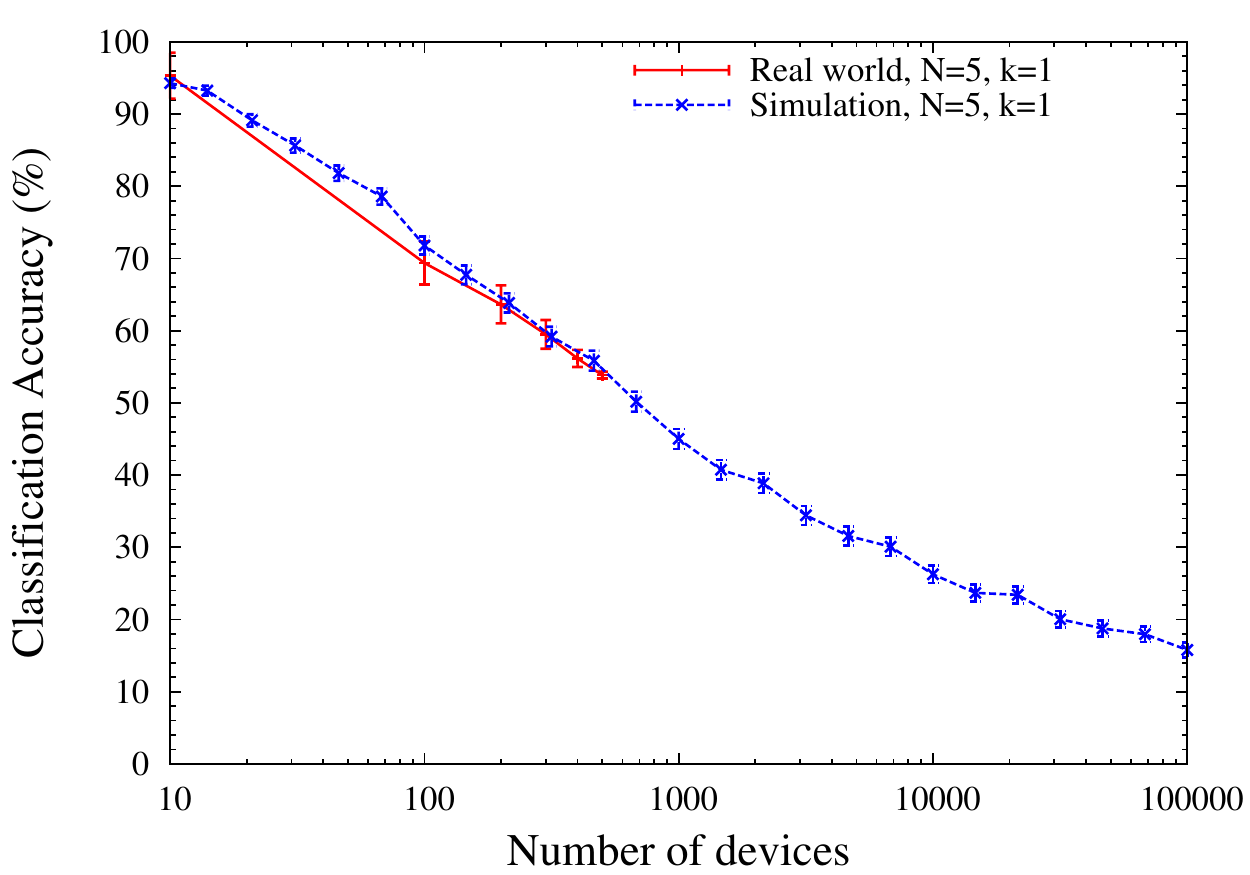,width=0.32\linewidth,clip=}
\end{tabular}
\caption{Comparison between real-world and simulation  .Simulation results closely match real-world results. Even with 100K devices we can} 
\label{knn_simulation}
\end{figure*}

Next, we run our \emph{k}-NN simulator for large number of devices. Figure~\ref{knn_simulation} shows the average classification accuracy achieved for different values of $N$ and $k$. Given that a user spends on average anywhere between 15 to 20 seconds on a web page~\cite{avgtimeonweb2,avgtimeonweb1} values of $N \leq 5$ seem most realistic (each of our data sample is 5 seconds worth of web session). We also experimented with other values of $k$, but found that setting $k=1$ provides the best overlap between real-world and simulation results.\footnote{%
Differences between our $k$-NN model and the actual $k$-NN classifier on real data arise from an imperfect fit of the distribution as well as the fact that our model makes an assumption that intra- and inter-phone distances are identically and independently distributed.} 
From figure~\ref{knn_simulation} we see that our simulation results closely match our real-world results. Also we can see that the average classification accuracy is in the range of 12-16\% when we scale up to 100\,000 devices. This accuracy is unlikely to be sufficient if motion sensors are the unique source of a fingerprint, but it suggests that combining motion sensor data with even a weak browser-based fingerprint is likely to be effective at distinguishing users in large populations.  Additionally, these classification accuracies are conservative and potentially provide a lower bound on performance, as random forests provide significantly better performance.

\section{Websites Accessing Sensors}{\label{measurement_study}}
In this section we look at how many of the top websites access motion sensors. We also try to cluster the access patterns into broad use cases. 

\begin{figure}[!h]
\centering
\includegraphics[width=1.0\columnwidth]{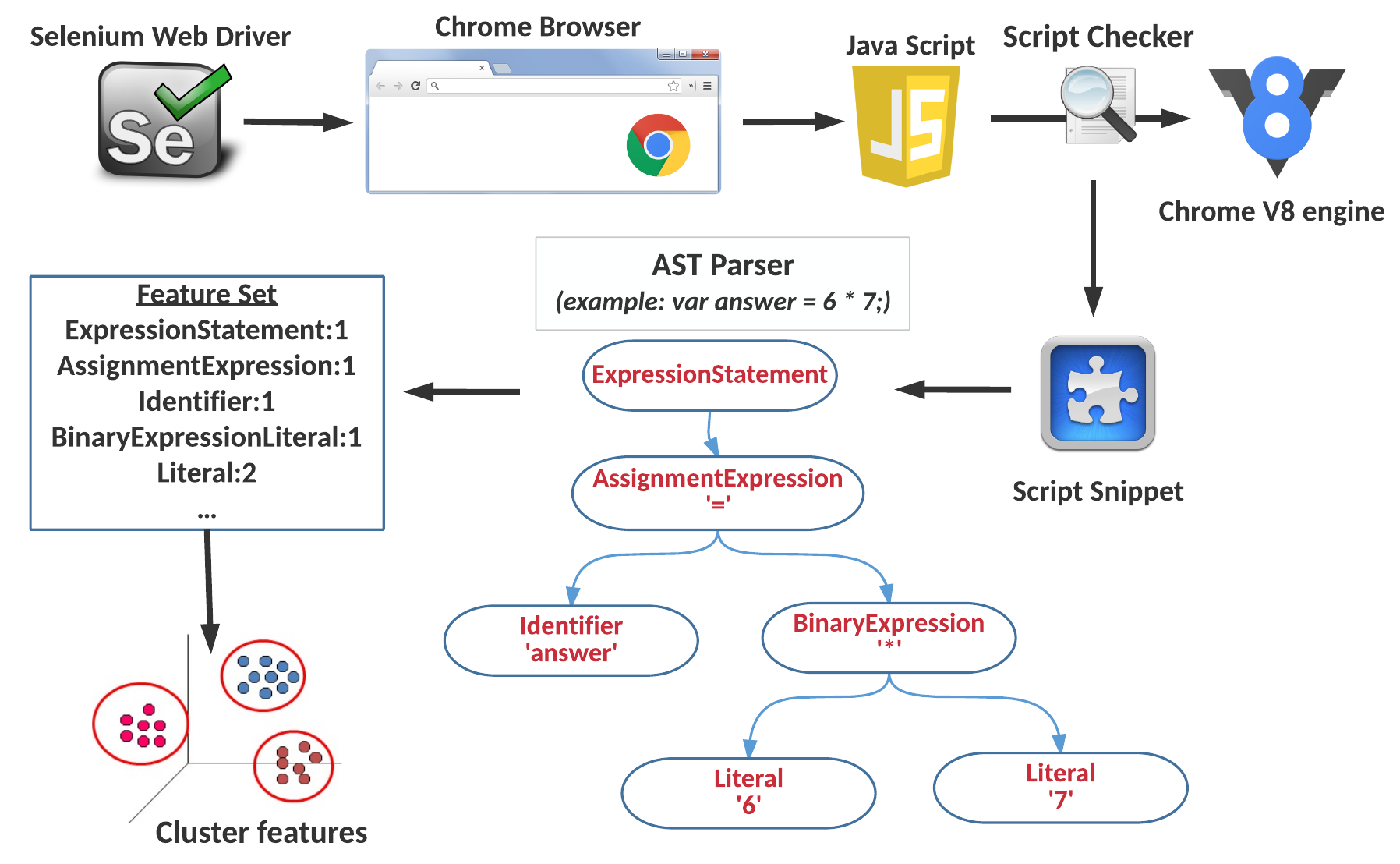}
\caption{Overview of our JavaScript analysis setup.}
\label{scan_JS} 
\end{figure}

\subsection{Methodology}  
Figure~\ref{scan_JS} provides an overview of our methodology to automatically capture and cluster JavaScripts accessing sensor data from mobile browsers. To automate this process, we use Selenium Web Driver~\cite{selenium} to run an instance of Chrome browser with a \emph{user agent} set for a smartphone client. In order to collect unfolded JavaScripts, we attach a debugger between the V8 JavaScript engine~\cite{v8_engine} and the web page. Specifically, we observe \texttt{script.parsed} function, which is invoked when new code is added with \texttt{<iframe>} or \texttt{<script>} tag. We implement the debugger as a Chrome extension and monitor all JavaScript snippets parsed on a web page. The debugger collects script snippets that access sensor data, i.e., scripts that invoke sensor APIs. Once scripts are collected, we aim to cluster them into a broad groups to identify their usage pattern. To analyze and quantify the similarity between JavaScript snippets, we parse them to produce \emph{Abstract Syntax Trees} (ASTs). ASTs have been used in prior literatures for JavaScript malware detection~\cite{Curtsinger:2011}. ASTs allow us to retain the structural and logical properties of the code while ignoring fine details like variable names, which are not useful for our analysis. We use the Esprima JavaScript parser~\cite{esprima} to visualize AST for each JavaScript snippet. We transform ASTs into normalized node sequences by performing \emph{pre-order traversal} on each tree. It should be mentioned that we start parsing each AST from the point where sensor data is first accessed. Each variable length sequence is composed of node types that appear in the tree. Since there are 88 distinct node types in JavaScript language, we transform the variable length normalized node sequences into 88-dimensional summary vectors. In other words, each JavaScript snippet is represented as a point in a 88-dimensional space, where each dimension corresponds to a node type. Finally, we attempt to perform unsupervised clustering on these summary vectors.

\subsection{Our Findings} 
We run our experiment for the top 100\,000 Alexa websites~\cite{Alexa}. Among these websites we find that 1130 websites contain some form of JavaScript code that accesses at least one of the motion sensors. It is worth mentioning that a few of the scripts were downloaded from \emph{ad networks} as the web pages were loaded. Table~\ref{realworld_internetscan} shows a breakdown of the detected websites into their corresponding ranking groups. We see that majority (1022 out of the 1130) of our detected websites come from the top 10\,000--100\,000 websites. However, even 6 of the top 100 websites seems to access motion sensors. 
 
\noindent\begin{minipage}{1.0\columnwidth}
\centering 
\captionof{table}{Top websites accessing motion sensors} 
\resizebox{0.4\columnwidth}{!}{
\begin{tabular}{|c|c|}
\hline
{Rank}&{\# of sites}\\
\hline
1--100&6\\
\hline
101--1000&12\\
\hline
1001--10000&90\\
\hline
10001--100000&1022\\
\hline
\end{tabular}}
\label{realworld_internetscan}
\end{minipage}

\begin{figure*}[!hb]
\noindent\begin{minipage}{1.0\linewidth}
\centering 
\captionof{table}{Generic use cases for accessing motion sensor data} 
\resizebox{1.0\columnwidth}{!}{
\begin{tabular}{|c|c|c|c|}
\hline  
{Cluster \#}&{\% of scripts}&{Use Case}&{Comment}\\
\hline
6&40.5&Transmit sensor data&Periodically sends motion sensor data to third party sites (can be marked suspicious)\\   
\hline
4&16.6&Random number generator&Crypto libraries use sensor data to add entropy to random numbers~\cite{sjcl}\\  
\hline
8&9.7&Detect device orientation&Detects device orientation periodically to readjust components in the website\\
\hline
5&8.9&Parallax scrolling/viewing&Parallax Engine that reacts to the orientation of a smart device~\cite{parallax}\\
\hline
2&7.1&Gesture detections&A \emph{jQuery} plug-in for gesture events such as `pinch', `rotate', `swipe', `tap' and `shake'~\cite{jGestures}\\
\hline
1&7.0&Motion captcha&A \emph{jQuery} CAPTCHA plug-in based on the HTML5 Canvas element~\cite{MotionCAPTCHA}\\
\hline
3&6.0&Miscellaneous&We were not able to point the exact use case for this cluster.\\
\hline
7&4.2&Specific Ad generation&Checks to see if accelerometer is present so that certain ad URLs can be requested\\
\hline

\end{tabular}}
\label{usecases}
\end{minipage}
\end{figure*}

\begin{figure}[!h]
\centering
\includegraphics[width=0.7\columnwidth]{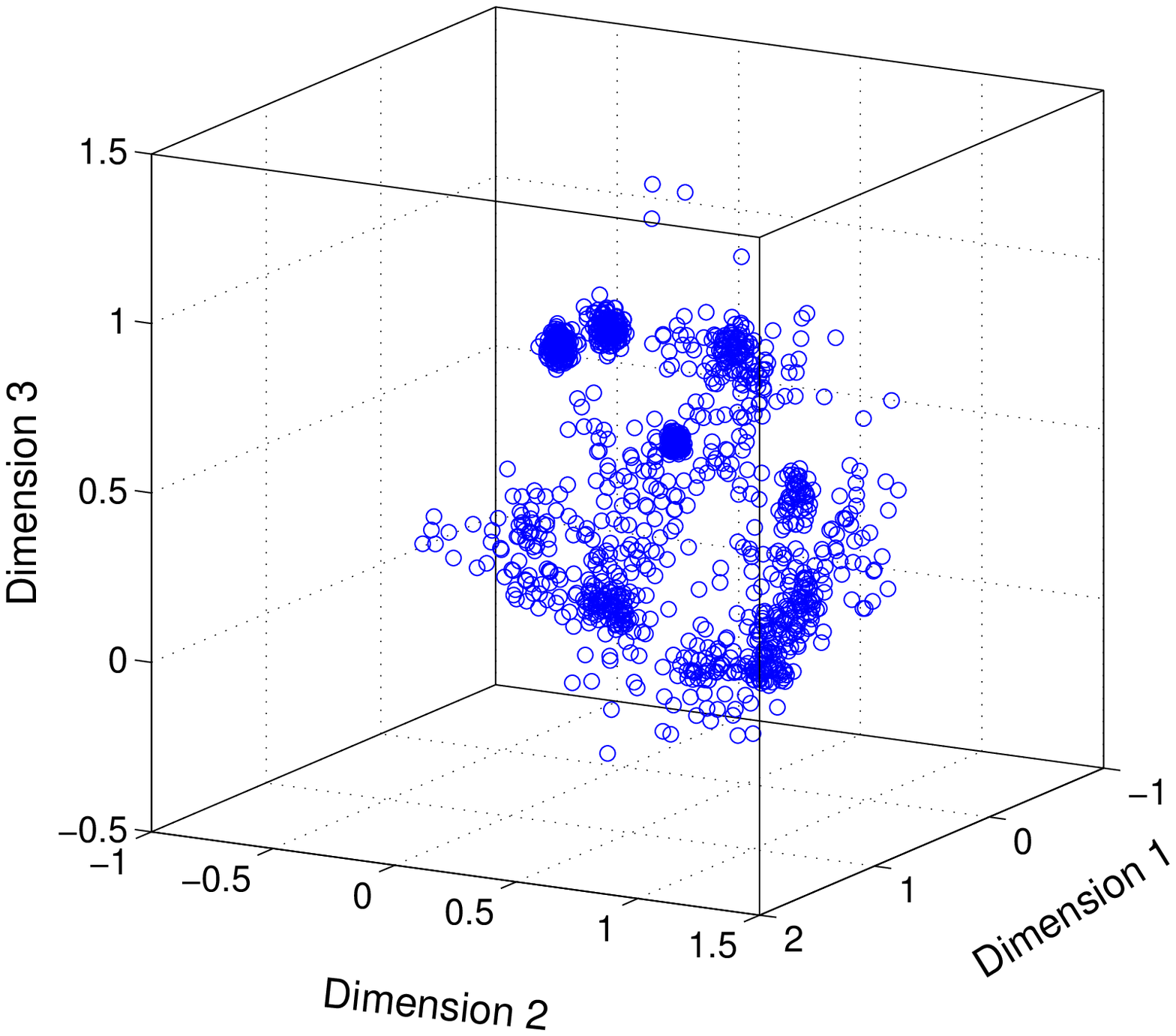}
\caption{Scatter plot for Javascript snippets accessing motion sensors along reduced dimensions.}
\label{scatterplot}
\end{figure}

\begin{figure}[!h]
\centering
\includegraphics[width=0.7\columnwidth]{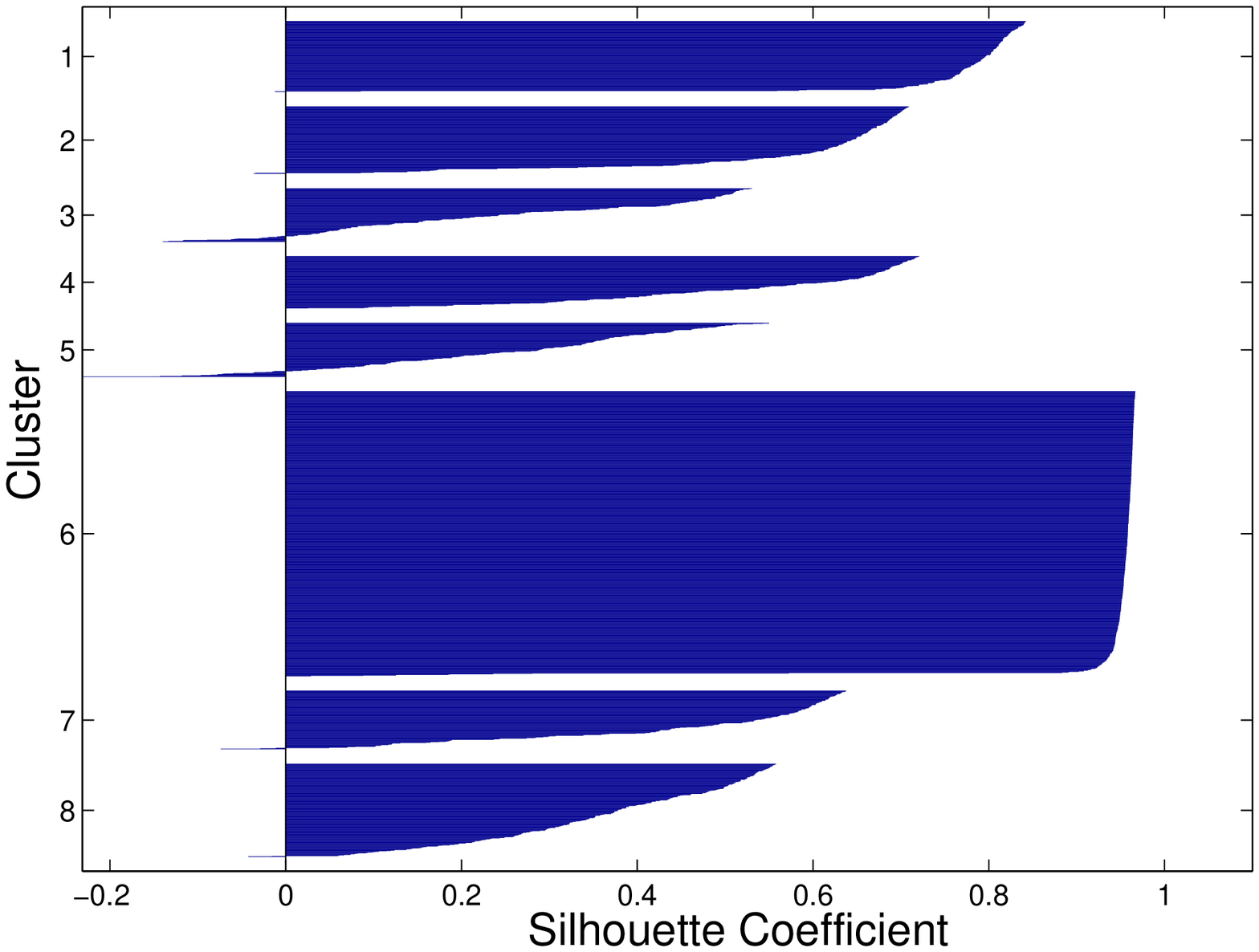}
\caption{Silhouette plot for the estimated 8 clusters. Clusters 1,2,4,6 and 7 have silhouette coefficient value greater than 0.6.}
\label{clusters}
\end{figure}

Our next goal is to cluster these 1130 websites into individual groups based on their usage of sensor data, so that we can identify the major reasons as to why websites access motion sensors. To cluster the JavaScript snippets into a small number of groups we first perform feature reduction to remove irrelevant features. Many of the 88 features had a value of zero for all Javascript snippets, so we first throw out these features. This reduces the size of the feature vector to 31. We then use the MATLAB Toolbox provided by Laurens van der Maaten~\cite{dim_reduction} to further map the features into a low dimensional space. We find that \emph{Stochastic Proximity Embedding} (SPE) method~\cite{agrafiotis2003stochastic} provides the best outcome in terms of both reducing dimensions and providing good clusters. Our final reduced feature space had three dimensions. Figure~\ref{scatterplot} shows a scatter plot along the three dimensions for all the JavaScripts. We can clearly see that the JavaScripts form clusters. To determine the number of clusters that is a good fit for our data we run \emph{k}-means clustering algorithm for different number of clusters and perform \emph{Silhouette} analysis~\cite{ROUSSEEUW198753}. Silhouette analysis can be used to study the separation distance between the resulting clusters. Silhouette coefficient ranges from +1, indicating point are very distant from neighboring clusters, through 0, indicating points are very close to the decision boundary between two neighboring clusters, to -1, indicating points are probably assigned to the wrong cluster. Table~\ref{silhouette_coeff} summarizes the average silhouette coefficients ($C_{silhouette}$) for different number of clusters. We see that silhouette coefficient peaks for 8 clusters. The corresponding silhouette plot for 8 clusters in given in figure~\ref{clusters}. We see that on average samples in cluster 1,2,4,6 and 7 have silhouette coefficient value greater than 0.6 while the samples in cluster 3,5 and 8 have silhouette coefficient close to 0.5. We also see some samples with negative silhouette coefficients and this is likely caused by JavaScripts coping code snippets belonging to two different libraries. Here, our goal is not to generate a perfect clustering of all the JavaScripts rather to broadly cluster them to identify the major usage patterns for accessing motion sensors.

\noindent\begin{minipage}{1.0\columnwidth}
\centering 
\captionof{table}{Silhouette coefficient for different number of clusters} 
\resizebox{1.0\columnwidth}{!}{
\begin{tabular}{|c|c|c|c|c|c|c|c|c|}
\hline
{Clusters}&3&4&5&6&7&8&9&10\\
\hline
{$C_{silhouette}$}&0.51&0.59&0.59&0.62&0.63&0.65&0.64&0.38\\
\hline
\end{tabular}}
\label{silhouette_coeff}
\end{minipage}

Once we have the general clusters we then go back to the JavaScripts to understand their usage for motion sensor data. This part of the analysis was carried out manually. However, since we generated 8 clusters we sampled multiple JavaScripts from each cluster to verify if they were performing similar functionality with the sensor data. We were able to identify 8 generic use cases for the motion sensors.

Table~\ref{usecases} summarizes our findings. We see that majority of the detected scripts periodically send sensor data to some third party sites. We were not able to pinpoint the exact usage for sending motion sensor data to third party sites as we did not have access to third party code. The next big usage for motion sensor data is that they are used in generating random numbers. Other uses cases include parallax viewing, gesture detection, motion captcha, specific ad generation and orientation detection. We were not able to concretely identify the use case for cluster 3 as we found that it contains multiple scripts all performing different tasks; some were doing touch analytics using accelerometer to detect tilt while others were doing something similar to parallax scrolling. We intend to perform a more thorough in-depth analysis of this usage patterns in the future.

\section{Countermeasures}{\label{sec:countermeasures}}
In this section we look at the performance and usability of two countermeasures against sensor-based smartphone fingerprinting. We evaluate sensor \emph{obfuscation}, one of the countermeasures proposed by Das et al.~\cite{anupam:2016}) and sensor \emph{quantization}, a new approach that we propose in this paper. We first look at their effectiveness against sensor fingerprinting. Next, we look are how these countermeasures impact the utility of the sensors by developing a web based \emph{labyrinth} game. .

\subsection{Obfuscation Vs. Quantization}
\label{sec:quantization}
First, we will briefly describe the operations of the countermeasures. Intuitively, obfuscation tries to randomize the sensor fingerprint by scattering the fingerprint at different locations in the feature space. On the other hand, quantization tries to group multiple fingerprints into the same location and thereby making it hard for the adversary to pinpoint the true device. The formal definition of the two approaches is given below.

\paragraphb{Obfuscation:} 
Obfuscation technique adds small amount of noise to the raw sensor values. The main idea is that since sensors themselves are not well calibrated, adding small noise to their raw value is equivalent to switching to a different (mis)calibrated sensor. We add obfuscation noise to the sensor data in the following manner: $s^O = s^M*g^O+o^O$, where $g^O$ and $o^O$ are the obfuscation gain and offset, respectively. Based from the study conducted by Das et al.~\cite{anupam:2016}, we set our offset and gain range to [-1.5,1.5] and [0.75,1.25], respectively.

\paragraphb{Quantization:} 
The basic idea behind quantization is that human brain cannot discriminate minute changes in angle or magnitude. As a result if the raw values of a sensor are altered slightly, it should not adversely impact the functionality of the sensor. We perform quantization in the polar coordinate system as it is easy to perceive. So, our first task is to covert the accelerometer data into its equivalent polar vector form as shown below:\nolinebreak
\begin{align}
radius, r&=\sqrt{a_x^2+a_y^2+a_z^2}\nonumber\\
inclination, \theta&=\cos^{-1} \frac{a_z}{r}\nonumber\\
azimuth, \psi&=\tan^{-1}\frac{a_y}{a_x}\nonumber
\end{align}
where $<a_x,a_y,a_z>$ represent the accelerometer data in the Cartesian coordinate system. Since gyroscope provides rotational rate in $rads^{-1}$, we do not perform any conversion for gyroscope data. Next we pass our sensor data through the following \emph{quantization} function:  
\begin{verbatim}
function quatization(val,type,bin_size){
  // val: raw sensor value
  // type: data type (angle or magnitude)
  // bin_size: quantization size 
  bin_num = floor(val/bin_size);
  remainder = mod(val,bin_size);
  if remainder >= binsize/2{
    bin_num = bin_num +1;
  }	
  return bin_num*bin_size;
}
\end{verbatim}
For angle related data ($\theta$,$\psi$ and gyroscope data) we set $bin_{size}=6$ while for magnitude (i.e., radius) we set $bin_{size}=1$. In other words, we place angles into 6 degree bins and for accelerometer magnitude we map it to the nearest integer. Once performing quantization on the accelerometer data, we remap it to Cartesian coordinate system using the following equations: 
$a_x=r\sin\theta\cos\psi$, $a_y=r\sin\theta\sin\psi$ and $a_z=r\cos\theta$.

\subsection{Effectiveness of Countermeasures}
In this section we will look at how the countermeasures impact the fingerprinting F-score. For this setup we run our fingerprinting scheme under three setting: baseline, obfuscation and quantization. For each setting we then evaluate F-score for both random forest and \emph{k}-NN (with LDML). Table~\ref{scheme_compare} shows our results for devices with at least 3 training samples (total 545 devices).

\noindent\begin{minipage}{1.0\columnwidth}
\centering 
\captionof{table}{Comparing obfuscation and quantization with baseline} 
\resizebox{0.8\columnwidth}{!}{
\begin{tabular}{|c|c|c|}
\hline
\multirow{2}{*}{Scheme}&\multicolumn{2}{c|}{Avg. F-score(\%)}\\
\cline{2-3}
&\emph{k}-NN with LDML&Random Forest\\ 
\hline
Baseline&50&78\\
\hline
Quantization&17&32\\
\hline
Obfuscation&7&26\\
\hline
\end{tabular}}
\label{scheme_compare}
\end{minipage}

\begin{figure}[!h]
\centering
\includegraphics[width=0.7\columnwidth]{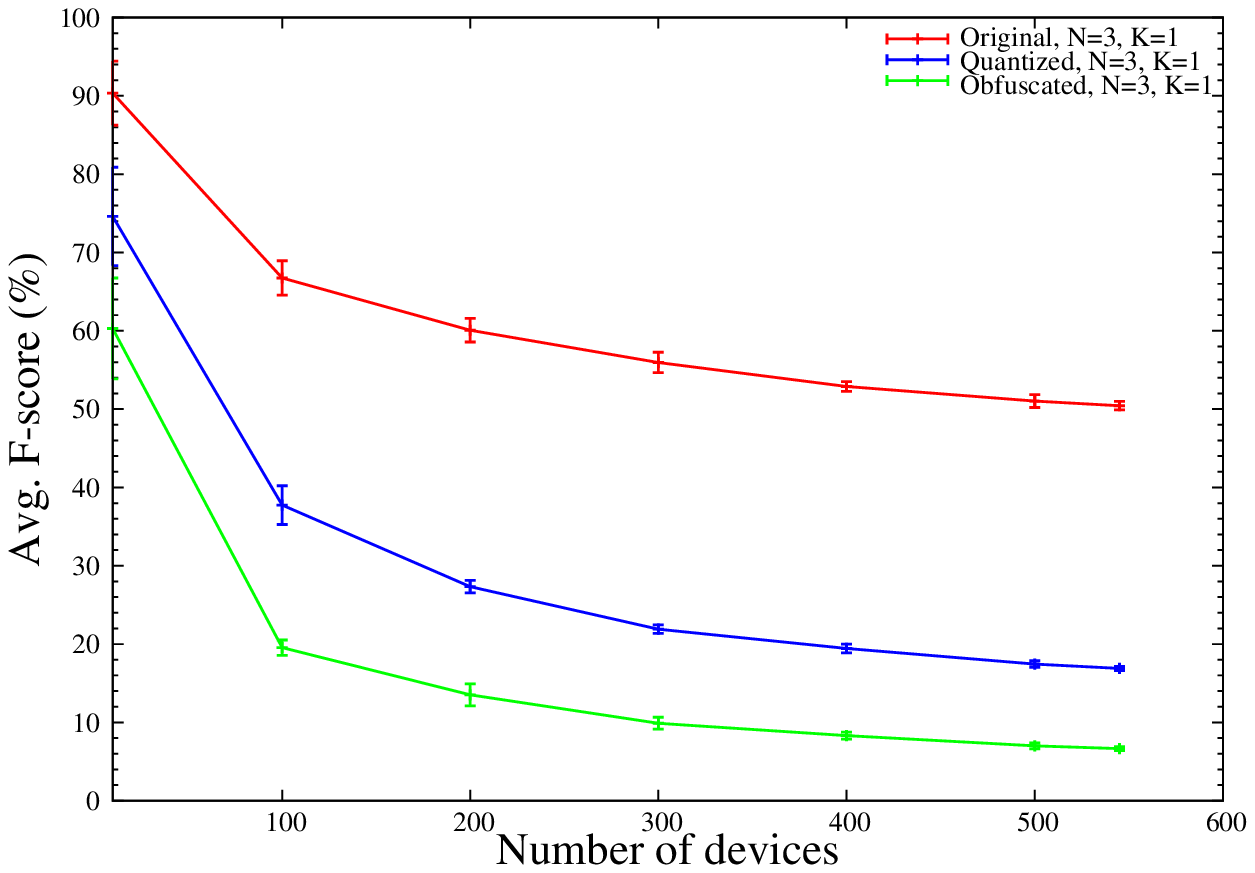}
\caption{Comparing countermeasure schemes against baseline.}
\label{dev_scheme} 
\end{figure}

We can see that the countermeasure schemes significantly reduce the F-score. Next we see how the countermeasure schemes react to different numbers of devices. Figure~\ref{dev_scheme} highlights our findings. We see that irrespective of the device number the F-score reduces significantly under both countermeasure schemes. Theses results indicate that simple countermeasures can thwart device fingerprinting significantly.

\subsection{User Study}
\label{sec:user_study}
The above countermeasures degrade the readings from the motion sensors and we wanted to better understand the impact of the countermeasures on the utility of the sensors to web applications. Of course, motion sensors have a wide range of uses, from simple orientation detection to activity classification, step counting, and other health metrics. Many of these, however, are deployed in application form, whereas we wanted to focus on the threat of fingerprinting by web pages. We performed a survey of web pages to identify how motions sensors are actually used. We found that one of the most common application of motion sensors was to detect orientation change in order to adjust page layout; such a drastic change in the gravity vector will be minimally impacted by countermeasures. We did, however, find several instances where web pages used the motion sensors as a means of gesture recognition in the form of tilt-based input controlling a video game. 

To study the impact of countermeasures on the utility of such tilt-based controls, we carried out a user study where participants were asked to play a game using tilt control while we applied privacy countermeasures to their motion sensor data. We then evaluated the impact of the countermeasures through both objective metrics of in-game performance, as well as subjective ratings given by the participants. Our study was approved by our institutional research board (IRB).

\subsubsection{Study Design} 
\begin{figure}[h]
\subfigure[Level 2 of the game]{
\centering
\includegraphics[width=0.48\linewidth]{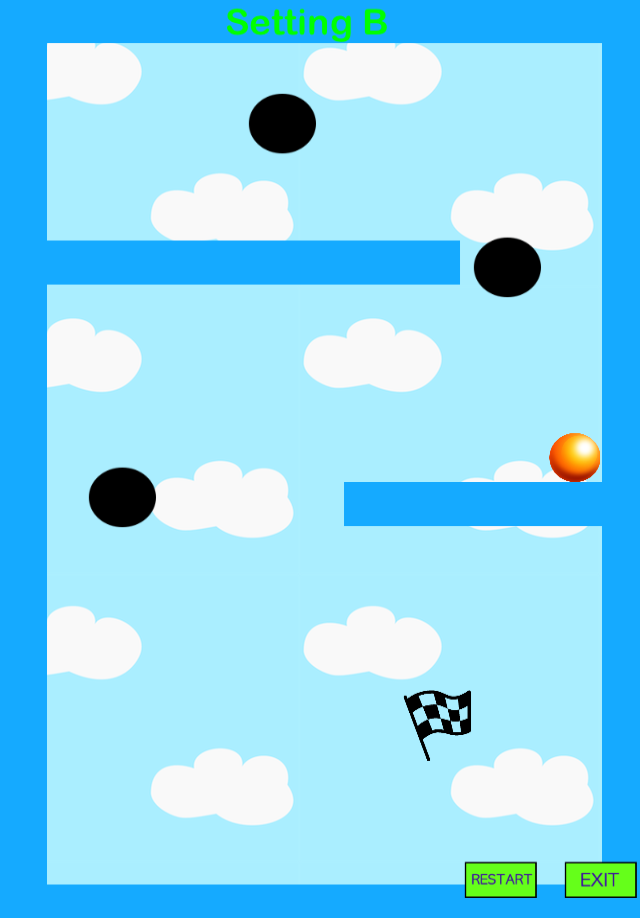}
\label{fig:gameplay}}
\subfigure[Feedback form]{
\includegraphics[width=0.48\linewidth]{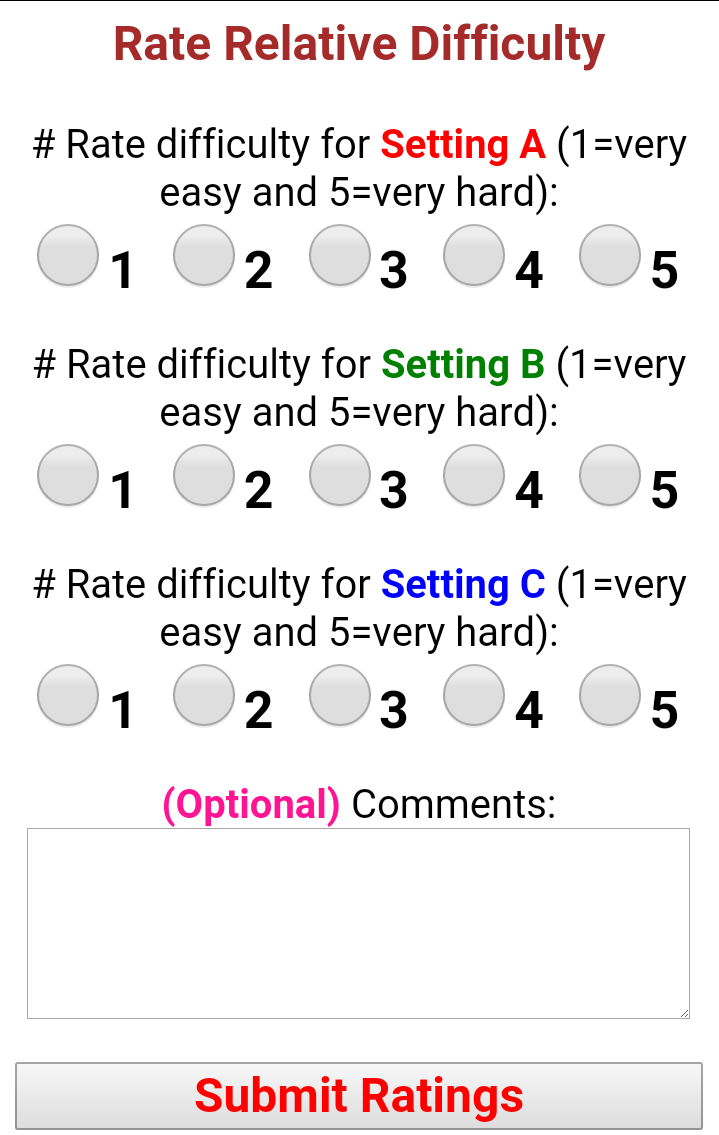}
\label{fig:feedback}}
\caption{Game interface. The object is to roll the ball to the flag while avoiding traps by tilting the smartphone. The user
is then asked for feedback about the relative difficulty of each level using different privacy settings.}
\end{figure}

After receiving some information about the study, our participants were invited to play a game using their personal smartphone (Figure~\ref{fig:gameplay}). The objective of the game is to roll a ball to its destination through a maze, while avoiding traps (hitting a trap restarts the level from the beginning). The game had five levels, which the participants played in order of increasing difficulty. Each level was played three times with different privacy countermeasures applied: baseline (no countermeasures), obfuscation, and quantization. The order of countermeasure settings was randomized for each participant and for each level, and not revealed to the participants. After completing a level three times, the participants were asked to rate each of the three settings in terms of difficulty of controlling the game on a scale of 1 to 5. Participants also were invited to provide free-form feedback (Figure~\ref{fig:feedback}).
Their ratings and feedback, along with the settings and metrics regarding the time spent on each game, and the number of times the game was restarted due to traps, were then sent to our server for analysis.

After completing a level, a user is invited to play the next level. Users were required to play levels in order of increasing difficulty, but participants were allowed to replay previous levels. We identified such repeat plays by setting a cookie in a user's browser and discarded repeat plays in our analysis.

\begin{table}[!h]
\begin{center}
\caption{Number of users that completed the first $n$ levels recruited through Amazon's Mechanical Turk and other means.}
\label{level-table}
\resizebox{0.7\columnwidth}{!}{
\begin{tabular}{|c|c|c|c|}
\hline {\bf Levels} & {\bf MTurk} & {\bf non-MTurk} & Total \\
{\bf completed} & & & \\ \hline \hline
1 & 0 & 26 & 26 \\ \hline
1--2 & 1 & 14 & 15 \\ \hline
1--3 & 0 & 34 & 34 \\ \hline
1--4 & 91 & 67 & 158 \\ \hline
1--5 & 107 & 63 & 170 \\ \hline \hline
{\bf Total} & 199 & 204 & 403 \\ \hline
\end{tabular}}
\end{center}
\end{table}

\begin{figure*}[h]
\includegraphics[width=\textwidth]{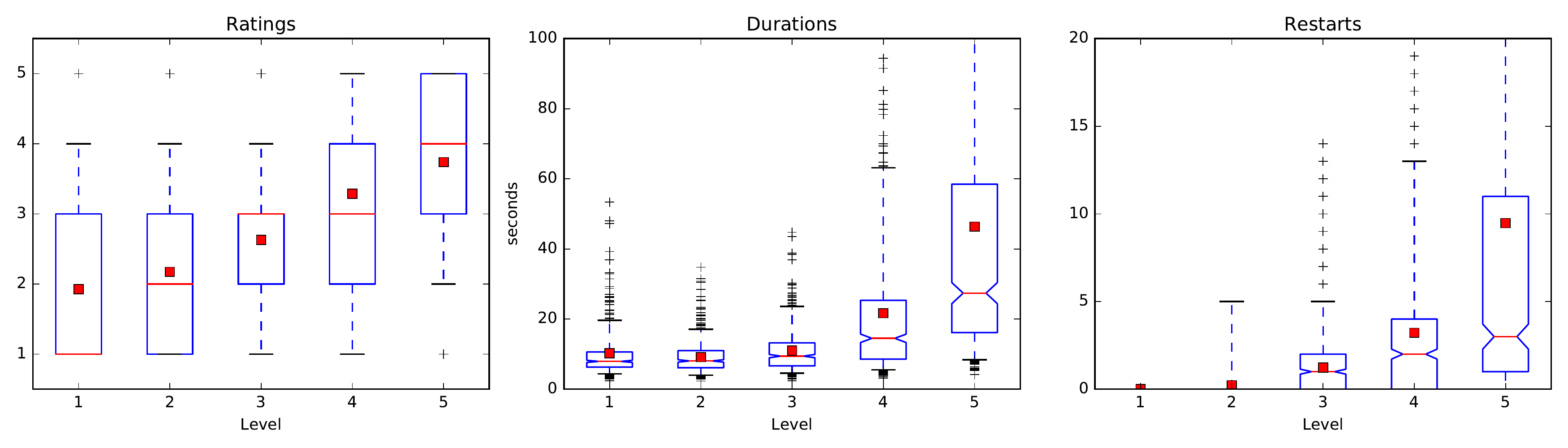}
\caption{Subjective and objective difficulty metrics increase across game levels. Box plots show the median (horizontal line) and the interquartile range (box), while whiskers show the range between the 5th and 95th percentile, with the outliers being individually represented. The notch in the box denotes the 95\% confidence interval around the median. Note: level 1 has no traps.}
\label{fig:difficulty-per-level}
\end{figure*}

\begin{figure*}[h]
\includegraphics[width=\textwidth]{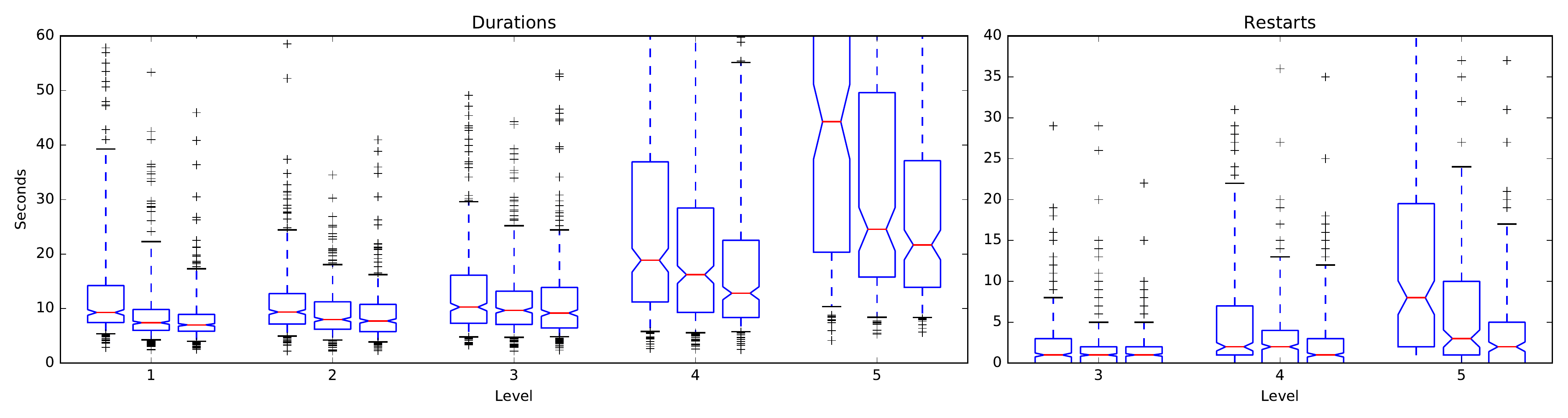}
\caption{Reduced game durations and number of restarts, as each level is played three times. A large training effect is observed between the first and second attempt, with a smaller effect between the second and third. Restarts on levels 1 and 2 are not shown.}
\label{fig:training-effect}
\end{figure*}

\begin{figure*}[!h]
\includegraphics[width=\textwidth]{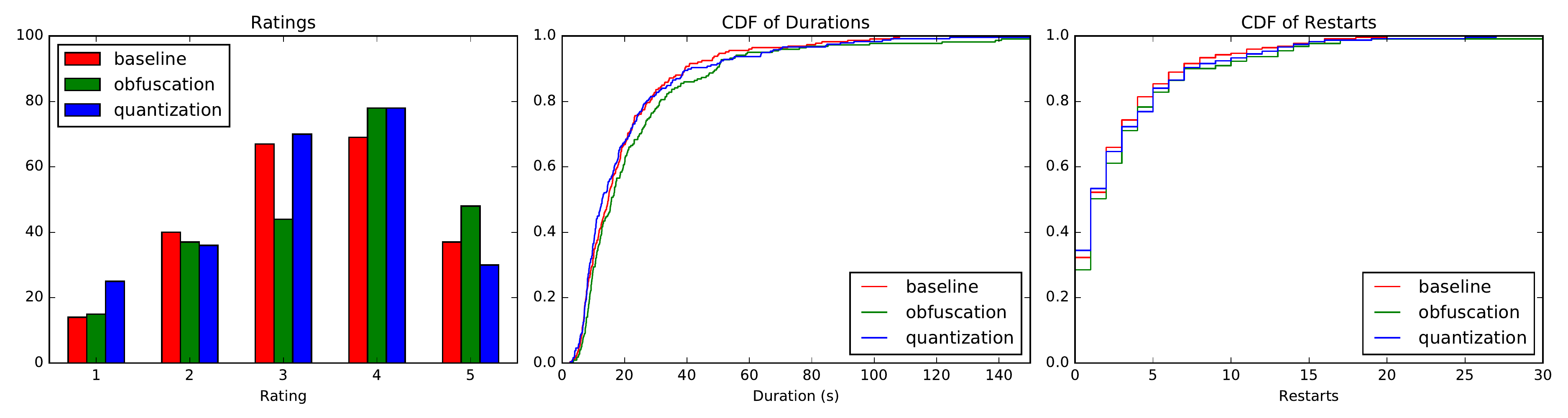}
\caption{Impact of privacy method on subjective and objective ratings, when considering second and third attempts only. Shown are the histogram of subjective ratings and CDFs of game durations and number of restarts on level 3. No significant difference is observed in any of the metrics.}
\label{fig:level-3}
\end{figure*}

\subsubsection{Study Results} We recruited users through institutional mailing lists, social media, as well as Amazon's Mechanical Turk. We collected data from 202 users via Mechanical Turk and 206 users that were recruited through other means, for a total of 408 users; several users' data had to be discarded due to irregularities in data collection. Note that not all users played through all five levels, as shown in Table~\ref{level-table}. Note that Mechanical Turk users had to complete five levels to receive their reward, but in some cases we were not able to receive some of their data due to network congestion at our server.

We found that, when considering the entire data set, the choice of privacy protection method did not significantly influence the subjective ratings assigned to the level ($\chi^2$ test, $p = 0.34$) nor the objective metrics of the game duration (pairwise t-tests, $p=0.10$ and $0.75$ comparing baseline to obfuscation and quantization, respectively) or the number of restarts due to traps (pairwise t-tests, $p=0.11$ and $0.47$). However, as expected, all difficulty metrics were significantly impacted by which level the person was playing, as shown in Figure~\ref{fig:difficulty-per-level}.

Furthermore, we observed a significant training effect between the first and second time a user played the level (each level is played a total of 3 times using different privacy methods), as seen in Figure~\ref{fig:training-effect}. Interestingly, this was not reflected in the subjective ratings (as verified by a $\chi^2$ test for each level), suggesting that participants corrected for the training effect during their reporting. There was a smaller training effect between the second and third time a level was played; the improved performance was statistically significant only for durations of levels 4 and 5 and for the number of restarts on level 5; which makes sense given the difficulty of these levels.  

We therefore compared the difficulty of metrics for different privacy methods across only the second and third attempts at a level, discarding the first attempt as training. For reasons of space, we show the results for level 3 only in Figure~\ref{fig:level-3}. Results for other levels are similar. Significance tests fail to detect any differences between the difficulty metrics when privacy methods are applied on any level.\footnote{The raw $p$-value comparing the number of restarts on level 5 between baseline and obfuscated cases is $0.025$ but note that this is not significant at a $p<0.05$ level after the Bonferroni correction is applied.}

\paragraphb{Limitations:} Although the study failed to detect a significant impact of privacy methods on utility, it does not definitively show that no impact exists---failure to reject a null hypothesis does not demonstrate that the null hypothesis is true. In particular, given the large variance in game performance across users, as seen in, e.g., Figure~\ref{fig:difficulty-per-level}, we would like to compare how different privacy methods change a single user's performance; however, given the low impact of privacy protection we have observed so far, we would need to modify our study to reduce or eliminate the training effect. Additionally, we tested our privacy methods in a short game, and perhaps in games with a longer duration some effects would materialize. However, we feel our results are promising in showing that users may not have to lose much utility to employ privacy protection methods.

\section{Conclusion}{\label{conclusion}}

We demonstrated that sensor fingerprinting is feasible on a much larger scale than previously studied. We showed that 90\% accuracy can be achieved for up to 400 devices, and \emph{at least} 12--16\% accuracy can be realized with 100\,000 devices, as predicted according to our model. Our measurement study reveals that motion sensors are already used by over 1\% of the top 100\,000 websites, and that sensor data are often sent to servers, which could serve as a vehicle for fingerprinting. Thus we can conclude that motion sensor fingerprinting is a realistic threat to mobile users' privacy.

We also evaluated the tradeoff between privacy and utility as realized by two different fingerprinting mitigation strategies. Our measurement study suggests that many applications of sensor data are unlikely to be affected. Our user study shows that even for sensitive applications that use motion sensors as control input, there is no significant impact of privacy mitigation  techniques on the usability of motion sensors in this context, according to both subjective and objective metrics.

\footnotesize{
\bibliographystyle{abbrv}
\bibliography{bibliograph}  
}


\end{document}